\documentclass{aa}
\pdfoutput=1 %for arXiv submission 
\usepackage{graphicx}
\usepackage{txfonts}
\usepackage{lscape}
\usepackage{booktabs}
\usepackage{threeparttablex}
\usepackage{wasysym}
\usepackage{arydshln} 
\usepackage{mathrsfs}
\usepackage{multirow}
\usepackage{amssymb}
\usepackage{xcolor}
\usepackage{ulem}
\usepackage{float}
\usepackage{placeins}
\usepackage{hyperref}
\hypersetup{colorlinks=true,linkcolor=blue, filecolor=magenta,citecolor=blue,urlcolor=cyan}
\newcommand{\xmm}{{\textit{XMM-Newton}}\/}
\newcommand{\nicer}{{\textit{NICER}}\/}

\newcommand{\swift}{{\textit{Swift}}\/}
\newcommand{\cha}{{\textit{Chandra}}\/}
\newcommand{\hst}{{\textit{Hubble Space Telescope}}\/}
\newcommand\T{\rule{0pt}{2.4ex}}       % Table top spacing                                         
\newcommand\B{\rule[-1.0ex]{0pt}{0pt}} % Table bottom spacing
\begin{document}
\title{\textit{Eppur si muove:} Evidence of disc precession or a sub-milliparsec SMBH binary in the QPE-emitting galaxy GSN~069}

   \author{
   G. Miniutti\thanks{gminiutti@cab.inta-csic.es}\inst{1},
   A. Franchini \inst{2,3} \and
   M. Bonetti \inst{3,4} \and
   M. Giustini \inst{1} \and
   J. Chakraborty \inst{5} \and
   R. Arcodia\thanks{NASA Einstein Fellow} \inst{5} \and
   R. Saxton \inst{6} \and
   E. Quintin \inst{7} \and
   P. Kosec \inst{8} \and
   I. Linial \inst{9,10} \and
   A. Sesana \inst{3,4}
}
   \institute{Centro de Astrobiolog\'ia (CAB), CSIC-INTA, Camino Bajo del Castillo s/n, 28692 Villanueva de la Ca\~nada, Madrid, Spain 
    \and 
    Universität Zürich, Institut für Astrophysik, Winterthurerstrasse 190, CH-8057 Zürich, Switzerland
    \and
    INFN, Sezione di Milano-Bicocca, Piazza della Scienza 3, I-20126 Milano, Italy
    \and
    Universit\`a degli Studi di Milano-Bicocca, Piazza della Scienza 3, I-20126 Milano, Italy
    \and
    Kavli Institute for Astrophysics and Space Research, Massachusetts Institute of Technology (MIT), Cambridge, MA, USA
    \and
    Telespazio UK for the European Space Agency (ESA), European Space Astronomy Centre (ESAC), Camino Bajo del Castillo s/n, 28692 Villanueva de la Ca\~nada, Madrid, Spain
    \and
    European Space Agency (ESA), European Space Astronomy Centre (ESAC), Camino Bajo del Castillo s/n, 28692 Villanueva de la Ca\~nada, Madrid, Spain
    \and
    Center for Astrophysics | Harvard \& Smithsonian, 60 Garden Street, Cambridge, MA 02138, USA
    \and
    Department of Physics and Columbia Astrophysics Laboratory, Columbia University, New York, NY 10027, USA
    \and
    Institute for Advanced Study, 1 Einstein Drive, Princeton, NJ 08540, USA
}

\date{Received  / Accepted  }
\date{}

\abstract {
X-ray quasi-periodic eruptions (QPEs) are intense soft X-ray bursts from the nuclei of nearby low-mass galaxies typically lasting about one hour and repeating every few. Their physical origin remains debated, although so-called impacts models in which a secondary orbiting body pierces through the accretion disc around the primary supermassive black hole (SMBH) in an extreme mass-ratio inspiral (EMRI) system are considered promising. In this work, we study the QPE timing properties of GSN~069, the first galactic nucleus in which QPEs were identified, primarily focusing on Observed minus Calculated (O-C) diagrams. The O-C data in GSN~069 are consistent with a super-orbital modulation on tens of days whose properties do not comply with the impacts model. We suggest that rigid precession of a misaligned accretion disc or, alternatively, the presence of a second SMBH forming a sub-milliparsec binary with the inner EMRI is needed to reconcile the model with the data. In both cases, the quiescent accretion disc emission should also be modulated on similar timescales. Current X-ray monitoring indicates that this might be the case, although a longer baseline of higher-cadence observations is needed to confirm the tentative X-ray flux periodicity on firm statistical grounds. Future dedicated monitoring campaigns will be crucial to test the overall impacts plus modulation model in GSN~069, and to distinguish between the two proposed modulating scenarios. If our interpretation is correct, QPEs in GSN~069 represent the first electromagnetic detection of a short-period EMRI system in an external galaxy, opening the way to future multi-messenger astronomical observations. Moreover, QPEs encode unique information on SMBHs inner environments, and can be used to gain insights on the structure and dynamics of recently-formed accretion flows or, possibly, to infer the presence of tight SMBH binaries in galactic nuclei.
}

\titlerunning{Disc precession or a sub-milliparsec SMBH binary in GSN~069}
\authorrunning{Miniutti et al.}

\maketitle

\section{Introduction}
\label{sec:intro}

Extreme-variability events associated with supermassive black holes (SMBHs) in galactic nuclei are being detected with increasing frequency in recent years. As the baseline of follow-up observations continuously extends, recurring optical, UV, and X-ray events, sometimes showing signs of a periodic or quasi-periodic repetition pattern, have been revealed. Some of these repeating nuclear transients (RNTs) likely result from the partial stripping (or disruption) of a stellar orbiting companion at pericentre as suggested, to cite a few examples, for ASASSN-14ko - an optically-detected RNT with mean recurrence time of $\simeq 115$~d \citep{2021ApJ...910..125P,2022ApJ...926..142P,2024ApJ...974...80B}, eRASSt~J045650.3-203750 - an X-ray and UV RNT with a recurrence time that decayed from $\simeq 300$~d to $\simeq 190$~d  in about $30$ months \citep{2023A&A...669A..75L,2024A&A...683L..13L}, and Swift~J023017.0+283603 that exhibited recurrent X-ray bursts every $\simeq 22$~d for a period of few hundred days \citep{2023NatAs...7.1368E,2024NatAs...8..347G,2024arXiv241105948P}. On shorter timescales, fast, high-amplitude soft X-ray bursts known as X-ray quasi-periodic eruptions (QPEs) have been unveiled in the past few years from the nuclei of low-mass galaxies \citep[for the first detection, see][]{2019Natur.573..381M}. 

Whether the same or similar physical processes give rise to the variety of observed properties and timescales in the current population of quasi-periodic RNTs remains to be understood. On the other hand, most (if not all) of the cases mentioned above are likely associated with the interaction between the central SMBH (or its accretion flow) and orbiting secondary companions in extreme mass-ratio inspiral (EMRI) systems, possibly representing the electromagnetic signature of one promising class of sources of gravitational waves to be detected by future experiments \citep[e.g.][]{Luo16,2017arXiv170200786A,2017PhRvD..95j3012B,2021ExA....51.1333S}. 

In this work, we focus on the relatively new phenomenon of X-ray QPEs, first discovered in the nucleus of the galaxy GSN~069 \citep{2019Natur.573..381M}. X-ray QPEs are high-amplitude soft X-ray bursts repeating on timescales of hours to days that stand out with respect to an otherwise stable quiescent X-ray emission, likely from the accretion disc around a SMBH. Following their detection in GSN~069, QPEs have been identified in another eight galactic nuclei \citep{2020A&A...636L...2G,2021Natur.592..704A,2021ApJ...921L..40C,2023A&A...675A.152Q,2024A&A...684A..64A,2024arXiv240916908B,2024Natur.634..804N}. So far, QPEs are exclusively an X-ray phenomenon, and no counterpart has been detected in the radio, IR, optical, or UV (see \citealt{2024ApJ...963L...1L} for theoretical predictions regarding UV-QPE analogues). 

The X-ray spectral properties of QPEs in the different sources studied so far are remarkably similar. Their X-ray spectrum is thermal-like with typical temperature evolving from $kT\simeq 50$-$80$~eV to $\simeq 100$-$250$~eV and back during the event, and the spectral evolution during bursts suggests an expanding emitting region (assuming blackbody emission). QPEs have a duty cycle (QPE duration over out-of-QPE quiescence) of $10$-$30$\%, and their X-ray luminosity at peak is $10^{42}$-$10^{43}$~erg~s$^{-1}$, depending on the specific source. During each QPE, hysteresis is present in the $L_{\rm X}$-$kT$ plane, with a hotter rise than decay \citep[see e.g.][]{2022A&A...662A..49A,2023A&A...670A..93M,2023A&A...675A.152Q,2024ApJ...965...12C,2024arXiv240901938G,2024Natur.634..804N}. 

Nuclear optical spectra of QPE host galaxies exhibit narrow emission lines with properties indicating the presence of an ionising source in excess of pure stellar light \citep{2022A&A...659L...2W}. Despite being basically unobscured in the X-rays, none of the current QPE-emitting galaxies shows optical nor UV broad emission lines in their spectra. This rules out the presence of a currently active galactic nucleus, and supports the idea that the quiescent emission seen in the X-rays is associated with an accretion flow that is unable to sustain a mature broad line region, possibly being too compact. Velocity dispersion measurements from optical spectroscopy indicate black hole masses of $10^5-10^{6.5}$ in QPE galactic nuclei with the possible exception of an higher mass SMBH of $\simeq 1-7\times 10^7~M_\odot$ in eRO-QPE4 \citep{2024A&A...684A..64A}. 

Recently, a connection between QPEs and TDEs has emerged. The long-term evolution of the quiescent X-ray emission in GSN~069 is consistent with a long-lived TDE peaking around July 2010 and with a second, partial TDE 9~yr later, although that interpretation is likely not unique \citep{2023A&A...670A..93M}. GSN~069 also shows abnormal C/N ratio in its UV and X-ray spectra \citep{2021ApJ...920L..25S,2024arXiv240617105K} which is consistent with a TDE interpretation, possibly from an evolved or stripped star \citep{2024ApJ...973L...9M}. The recently-confirmed QPE source AT2019vcb \citep{2023A&A...675A.152Q,bykov2024}, or Tormund, was initially detected as an optical TDE \citep{2023ApJ...942....9H}, while XMMSL1~J024916.6-041244 \citep{2021ApJ...921L..40C} was classified, prior to QPE identification, as an X-ray-detected TDE \citep{2007A&A...462L..49E}. The X-ray decay of the quiescent emission in eRO-QPE3 is also suggestive of a TDE \citep{2024A&A...684A..64A}.  The recently reported QPEs in AT2019qiz occur in a well studied optically-selected TDE and were first detected about $4$~yr after optical peak \citep{2024Natur.634..804N}, confirming a clear connection between QPEs and TDEs as well as a likely delay in their appearance with respect to the TDE outburst, as noted already in GSN~069 by \citet{2019Natur.573..381M}. Finally, QPEs and TDEs appear to prefer the same type of low-mass, post-starburst host galaxies with a high incidence of extended narrow line regions  \citep{2024ApJ...970L..23W,2024ApJ...969L..17W}, pointing towards a scenario in which the nuclei of QPE galaxies were active in the past but have then switched off leaving only relic narrow emission lines, as suggested by \citet{2019Natur.573..381M} for GSN~069 whose optical spectrum is unambiguously that of a Seyfert~2 galaxy \citep{2013MNRAS.433.1764M,2022A&A...659L...2W}. This is also consistent with the analysis of \hst\ data of GSN~069 by \citet{2024MNRAS.530.5120P} who highlight the presence of a compact ($\lesssim 35$~pc) [O\,{\sc iii}] emitting-region likely ionised by the current, recently-activated emission. High-cadence X-ray monitoring of TDEs, especially at late times, may thus be key to discover new QPE-emitting galactic nuclei.

The physical origin of QPEs is still uncertain and the focus of active research. Several models have been proposed so far, and they cluster into two main scenarios: disc instability models \citep{2021ApJ...909...82R,2022ApJ...928L..18P,2023ApJ...952...32P,2023A&A...672A..19S,2023MNRAS.524.1269K}, and orbital models invoking the repeated interaction between the central SMBH and orbiting companions \citep[see e.g.][]{2020MNRAS.493L.120K,2022MNRAS.515.4344K,2021MNRAS.503.1703I,2021ApJ...917...43S,2022ApJ...926..101M,2022ApJ...941...24K,2022A&A...661A..55Z,2022ApJ...933..225W,2023MNRAS.524.6247L,2023ApJ...945...86L,2024A&A...682L..14W}. In this work, we focus on the QPE timing properties in  GSN~069, and we compare the timing behaviour to the theoretical predictions from one of the most popular QPE models, the so-called impacts model that invokes repeated collisions between an orbiting companion and the accretion disc around the primary SMBH in an EMRI system, with each collision giving rise to an X-ray  QPE \citep{2021ApJ...921L..32X,2023ApJ...957...34L,2023A&A...675A.100F,2023MNRAS.526...69T,2024PhRvD.109j3031Z,2024PhRvD.110h3019Z,2024arXiv240714578Y}.

After introducing a few relevant definitions in Section~\ref{sec:definitions}, we study the QPEs timing properties in GSN~069 in Section~\ref{sec:timing} and \ref{sec:OC}. The impacts model for QPEs is introduced in Section~\ref{sec:model}, where we discuss its predictions in comparison with the GSN~069 data highlighting a series of inconsistencies that can be cured by introducing an external modulation of QPEs arrival times. Two possible modulation scenarios are proposed in Section~\ref{sec:wayforward}, while Section~\ref{sec:BHmasses} discusses the current status on SMBH mass estimates in GSN~069. The two proposed modulation scenarios are compared with the data in Section~\ref{sec:discprecession} and \ref{sec:triples} by making use of numerical simulations of impact times between the secondary and the accretion disc around the primary SMBH. The quiescent (out-of-QPEs) X-ray flux variability from an X-ray monitoring campaign between May and September 2024 is studied in Section~\ref{sec:modulation}. We discuss our results and their implications in Section\ref{sec:discussion} and \ref{sec:conclusions}.

\section{QPE timing definitions}
\label{sec:definitions}

\begin{figure}[t]
\centering \includegraphics[width=0.9\columnwidth]{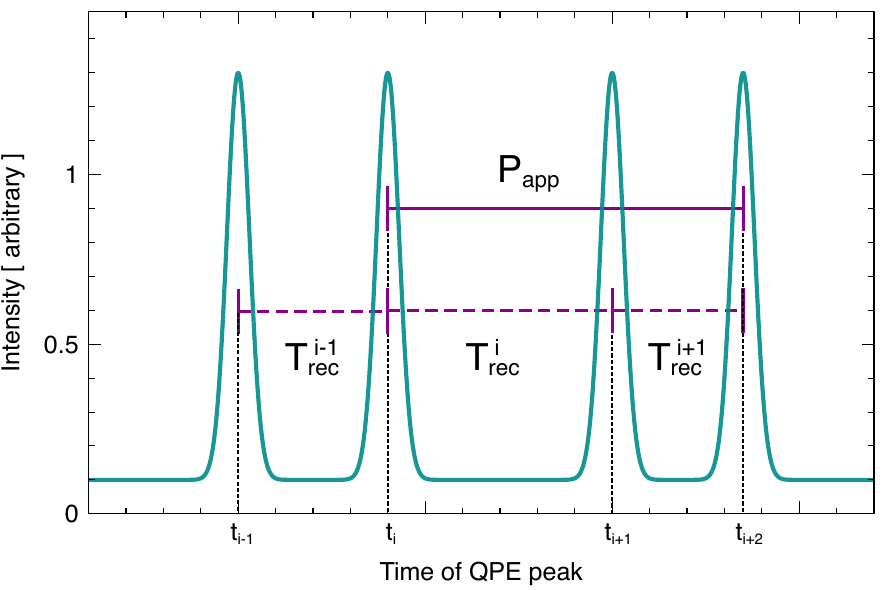}
%{\vspace{-0.2cm}}
\caption{Schematic representation of a QPE time series comprising four consecutive QPEs. Odd and even QPEs represent impacts through the ascending and descending nodes respectively (or vice-versa). The definition of the different QPE recurrence times that are used to compute $P_{\rm app}$ and $e_{\rm app}$ is highlighted (see Eq.~\ref{eq:T_rec}-\ref{eq:e_app} and text for details).}
\label{fig:Trec_def}
\end{figure}

Mostly based on the alternating longer and shorter recurrence times between consecutive QPEs in GSN~069, RX~J1301.9+2747, and eRO-QPE2, and on the presence of an X-ray quiescence consistent with accretion disc emission \citep{2019Natur.573..381M,2020A&A...636L...2G,2021Natur.592..704A}, models invoking twice-per-orbit collisions between a secondary object (a star or  black hole) and the accretion disc around the primary SMBH in an EMRI system have received considerable attention in the recent past. Here we introduce a few definitions that can be derived from QPEs peak times of arrival, and that are well suited to study the QPE timing properties within the context of the impacts model. As a note of caution, we point out that the intrinsic (unknown) delay between a collision and the corresponding QPE peak emission was here assumed not to vary across impacts. Under this assumption, the QPE arrival time is therefore taken as representative of the corresponding impact time. We stress that the amplitude of the delay itself is not relevant, as our analysis is based on differential quantities (e.g. the recurrence time between QPEs), but delays becomes potentially important if they vary across impacts. That delays are impacts-independent is not necessarily the case, so that all results presented in our work are likely subject to a certain degree of systematic error and the reported statistical uncertainties should be considered with some caution.

In Fig.~\ref{fig:Trec_def}, we show a schematic representation of a QPE time series comprising four consecutive QPEs where we define, for any QPE at time $t_i$, the QPE recurrence time $T_{\rm rec}$, the EMRI apparent orbital period ($P_{\rm app}$), and its apparent eccentricity ($e_{\rm app}$) as: 
\begin{align} 
T_{\rm rec}^{(i)} & = t_{i+1} - t_i \label{eq:T_rec}\\ 
P_{\rm app}^{(i)} & = T_{\rm rec}^{(i)} + T_{\rm rec}^{(i+1)} = t_{i+2} -t_{i}\label{eq:T_orb}\\ 
e_{\rm app}^{(i)} & = \left|T_{\rm rec}^{(i)} - T_{\rm rec}^{(i-1)} \right|~\left( T_{\rm rec}^{(i)} + T_{\rm rec}^{(i-1)} \right)^{-1}~. \label{eq:e_app}
\end{align}

In the context of the impacts model, odd and even QPEs are associated with collisions between the secondary EMRI component and the accretion disc around the primary at the ascending and descending nodes respectively  (or vice-versa). The time interval $P_{\rm app}$ between consecutive QPEs of the same parity thus provides an estimate of the EMRI orbital period $P_{\rm orb}$. Consecutive QPEs of different parity are instead separated by alternating longer and shorter $T_{\rm rec}$, unless the EMRI orbit is circular or the intersection between the disc and the orbital planes lies along the orbit's semi-major axis. By definition, $e_{\rm app}$ is an apparent EMRI eccentricity that, for Keplerian orbits, is constant and can take any value between zero and $e_{\rm app}^{\rm max}$, depending on the system geometry. The maximum ($e_{\rm app}^{\rm max}$) is reached when the difference between consecutive longer and shorter recurrence times is the largest, that is when the intersection between the orbital and disc planes is along the orbit's latus rectum. As a consequence, a measure of $e_{\rm app} \lesssim e_{\rm app}^{\rm max}$ represents a lower limit on the actual orbital eccentricity $e_{\rm orb}$.

\begin{figure}[t] 
    \centering
        \includegraphics[width=0.9\columnwidth]{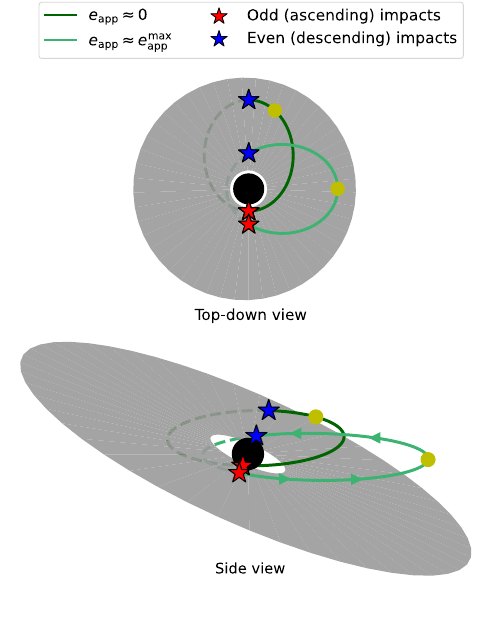}
    \caption{Effects of apsidal precession on secondary-disc impacts. We show two different apsidal phases leading to $e_{\rm app} \simeq e_{\rm app}^{\rm max}$, and to $e_{\rm app} \simeq 0$. The former is shown as lighter orbit and corresponds to an apsidal phase in which the difference between consecutive longer and shorter $T_{\rm rec}$ is maximal, with impacts occurring at approximately the same distance from the centre. Conversely, when $e_{\rm app} \simeq 0$, the two consecutive $T_{\rm rec}$ are approximately equal, and impacts occur close to apocentre and pericentre respectively. Ascending (descending) node's impacts are denoted as odd (even) impacts leading to odd (even) QPEs and are shown in red (blue).}
\label{fig:schema}
\end{figure}

In General Relativity, apsidal precession of the EMRI orbit implies that $T_{\rm rec}$, $P_{\rm app}$, and $e_{\rm app}$ vary periodically on the apsidal precession timescale. $P_{\rm app}$ is therefore not equal to the constant Keplerian orbital period, and $e_{\rm app}$ spans all possible values between zero and $e_{\rm app}^{\rm max}$ depending on precession phase. In fact, $e_{\rm app}\simeq0$ twice per apsidal period, that is when consecutive impacts occur close to apocentre and pericentre. A schematic view of the effects of apsidal precession on the impacts between the EMRI's secondary and the disc is shown in Fig.~\ref{fig:schema}. Note that, due to the different location of impacts on the disc with respect to the observer, light-travel-time delays have to be expected as well. Considering a typical EMRI nearly circular orbit with semi-major axis of $100~R_g$, and a SMBH mass of $10^6~M_\odot$, light-travel-time delays are expected not to exceed $10^3$~s, or $\simeq 17$~minutes. However, such relatively large delays are only obtained for a very specific geometry in which impacts align with the line-of-sight and occur close to apocentre and pericentre respectively. More generally, light-travel-time delays are not expected to exceed a few minutes. The impacts model predictions on all the quantities defined above  are discussed and compared with the observed QPE data of GSN~069 in Section~\ref{sec:model}. 

\section{Application to GSN~069}
\label{sec:timing}

\begin{figure*}[t]
\centering \includegraphics[width=1.8\columnwidth]{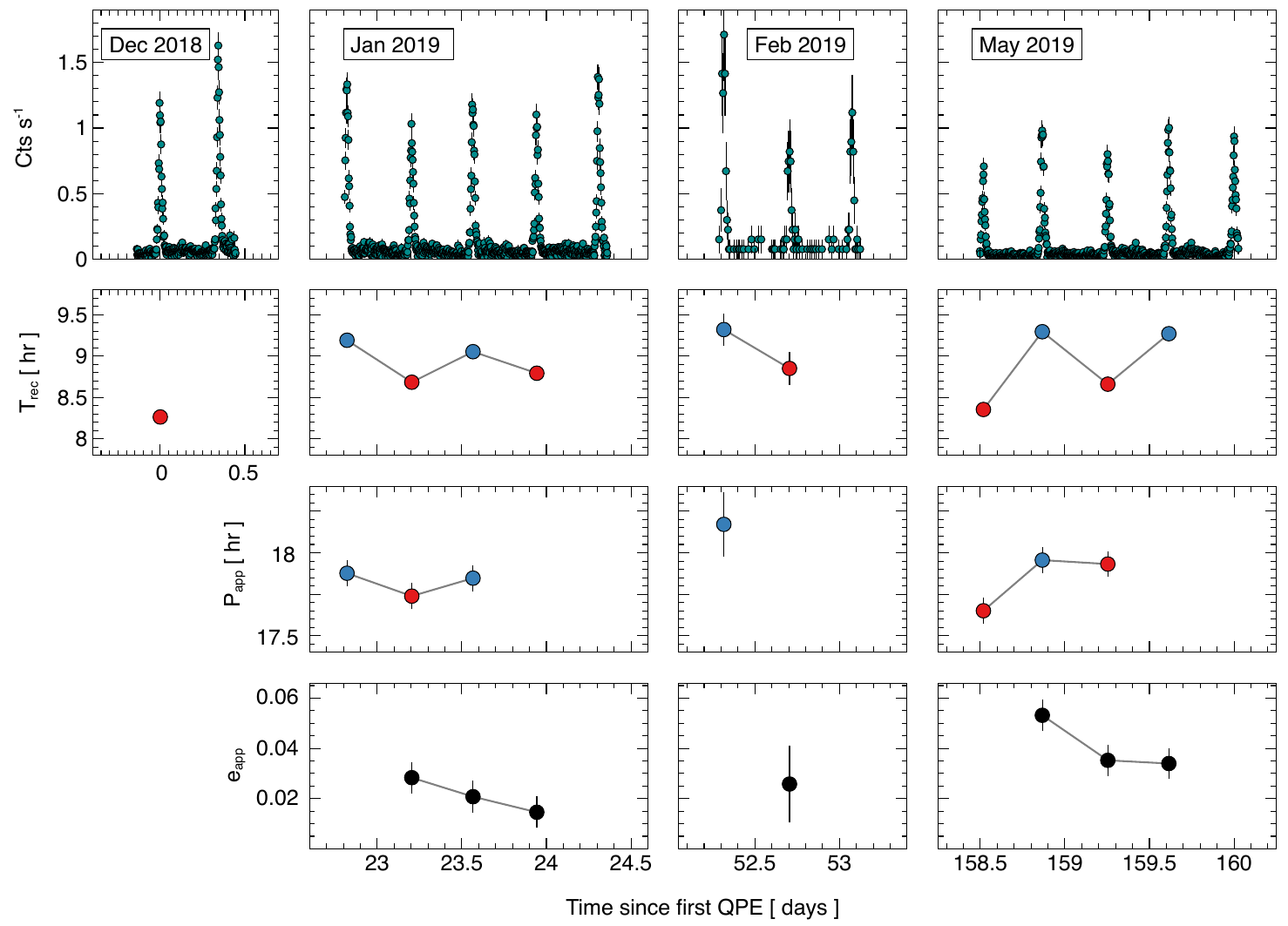}
%{\vspace{-0.2cm}}
\caption{QPE timing properties in GSN~069. In the upper row, we show the X-ray 0.4-1~keV light curves of GSN~069 used in this work. The first two and last light curves are from the EPIC-pn camera on board \xmm, while the third is from the ACIS-S detector on board \cha. The latter light curve has been re-scaled to the expected EPIC-pn count rate \citep[as in][]{2023A&A...670A..93M}. \xmm\ data could be used down to 0.2~keV but, since QPE properties (peak time and overall duration) are energy-dependent \citep{2019Natur.573..381M}, we use here a common energy band down to 0.4~keV only in order to make use also of the \cha\ data. The three lower rows show $T_{\rm rec}$, $P_{\rm app}$, and $e_{\rm app}$ as obtained from the QPE peak times. Due to the gaps in the data the colour code distinguishing odd from even QPEs during different observations might be different from the one shown here where we arbitrarily assigned all stronger (weaker) QPEs to the even (odd) time series.
}
\label{fig:gsn}
\end{figure*}

We considered QPE data of GSN~069 and we derived, from the observed X-ray light curves, the quantities defined in Eq.~\ref{eq:T_rec}-\ref{eq:e_app} from the observed QPEs peak times. Details on the observations used in this work and on relevant aspects of data analysis are given in Appendix~\ref{sec:data}. We make use here of the first four observations during which QPEs were detected, from December 2018 to May 2019. Soon after, the quiescent (out-of-QPEs) X-ray emission of GSN~069 experienced a sudden significant rebrightening, peaked for about $200$~d, and started to decay towards the previous X-ray flux level. QPEs were detected during the rise, disappeared at peak, and were detected again during the decay, but with timing properties significantly different from those preceding the rebrightening \citep{2023A&A...670A..93M,2023A&A...674L...1M}. As discussed by \citet{2023A&A...674L...1M}, the disappearance of QPEs at peak is not because they are overwhelmed by enhanced disc emission, but is rather associated with a significantly lower QPE peak temperature that approaches that of the quiescent disc emission. As an example, in observations with the highest quiescent level and no QPEs, the quiescent 0.2-1~keV count rate is $\simeq 0.7$~cts~s$^{-1}$, significantly lower than the typical QPE peak count rate (typically $\gtrsim 1.5$~cts~s$^{-1}$ in the same band), so that QPEs with similar amplitude and, most importantly, temperature as those that are observed at lower flux levels would have been easily detected. On the other hand, if the QPE peak temperature is similar to that of the underlying disc emission at high disc fluxes (or, better, at high mass accretion rates), QPEs would be observed as low-amplitude X-ray fluctuations since their bolometric luminosity is a relatively small fraction of the disc one \citep{2023A&A...670A..93M,2023A&A...674L...1M}. 

The X-ray rebrightening was consistent with a second partial TDE occurring about $9$~yr after the initial one in GSN~069, although this interpretation is only based on the X-ray light curve shape, and is thus likely not unique. Partial rather than full TDEs might also contribute to explain the long-lived nature of the X-ray emission in GSN~069 following its initial 2010 X-ray outburst \citep{2024ApJ...961L...2B}. The irregular QPE properties during the rebrightening rise and decay might then signal that the accretion flow was disturbed, re-arranging, and possibly rapidly precessing at that epoch \citep[see e.g.][]{2024ApJ...973..101L}, and could support QPE models in which the disc plays an important role, such as the impacts model. The properties of the QPEs detected after May 2019 are being studied in detail and will be presented in a forthcoming work, although some relevant results are anticipated in Section~\ref{sec:Pdotconfirm}. 

Fig.~\ref{fig:gsn} shows $T_{\rm rec}$, $P_{\rm app}$, and $e_{\rm app}$ for the four observations of GSN~069 considered here, together with the corresponding X-ray light curves. All quantities were derived from QPEs peak times of arrival obtained through constant plus Gaussian functions fits to the corresponding X-ray light curves, as briefly discussed in Appendix~\ref{sec:data}. The recurrence times between consecutive QPEs consistently alternate defining longer and shorter consecutive $T_{\rm rec}$, as expected in the impacts model for any non-zero EMRI eccentricity. We note that longer recurrence times always follow stronger QPEs, although this could be chance coincidence due to the limited number of detected QPEs (15). The separation between QPEs of the same parity ($P_{\rm app}$, a proxy for the EMRI's orbital period) is not constant. Finally, $e_{\rm app}$ is also variable with typical values suggesting that the EMRI eccentricity is low but non-zero in GSN~069. This is in agreement with results by \citet{2023A&A...675A.100F} and \cite{2024PhRvD.110h3019Z} who have applied different versions of the impacts model\footnote{While \citet{2023A&A...675A.100F} considered a precessing accretion disc, this was not included in the work by \cite{2024PhRvD.110h3019Z}} to GSN~069 deriving an EMRI's orbital period of $P_{\rm orb}\simeq 18$~hr and eccentricity of either $\simeq 0.1$ or $\lesssim 0.15$, both consistent with the average $P_{\rm app}$ and with $e_{\rm app}$ in Fig.~\ref{fig:gsn}.

\section{O-C analysis}
\label{sec:OC}

\begin{figure*}[t] 
    \centering
        \includegraphics[width=0.45\textwidth]{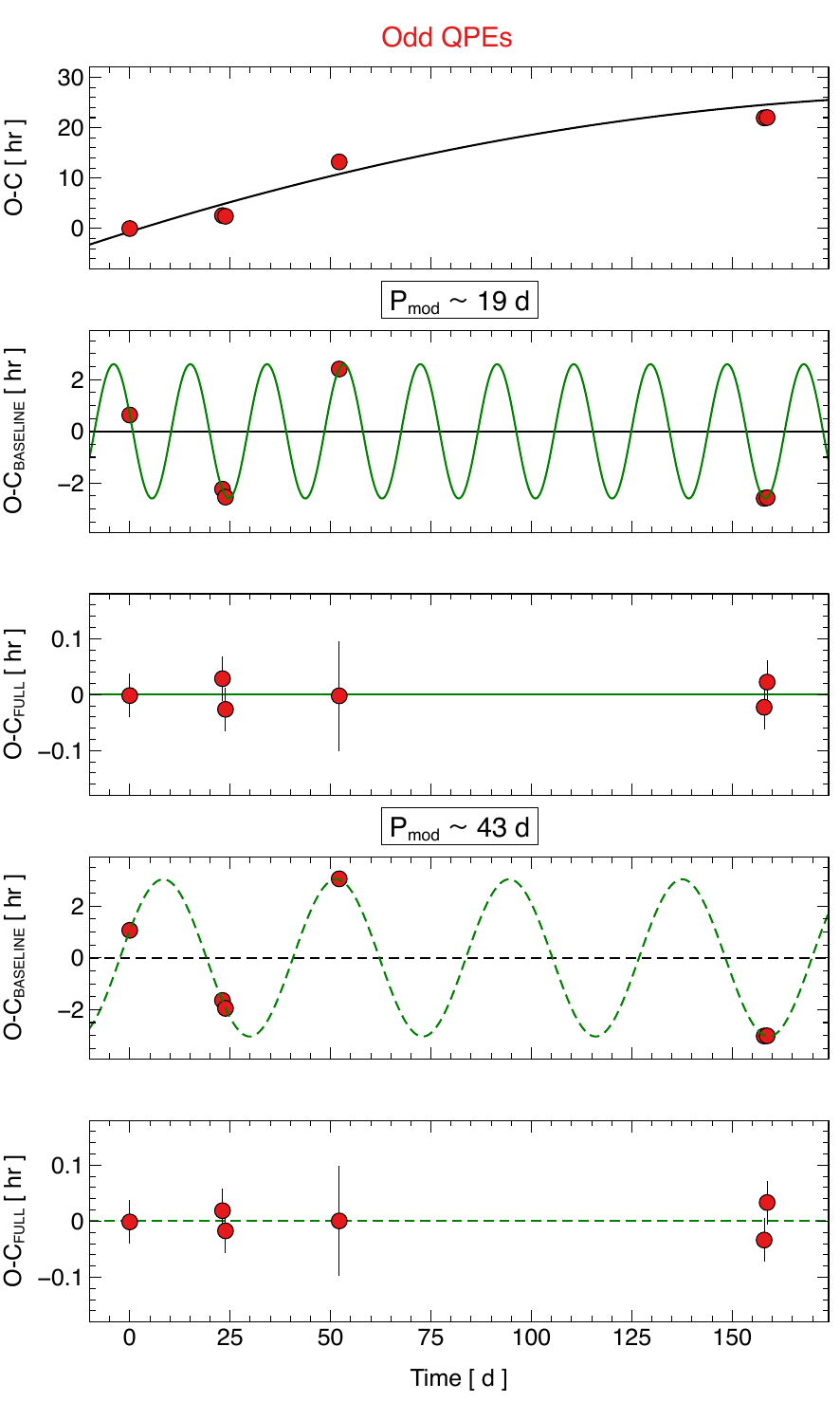}%
    \hspace{0.3cm}
        \includegraphics[width=0.45\textwidth]{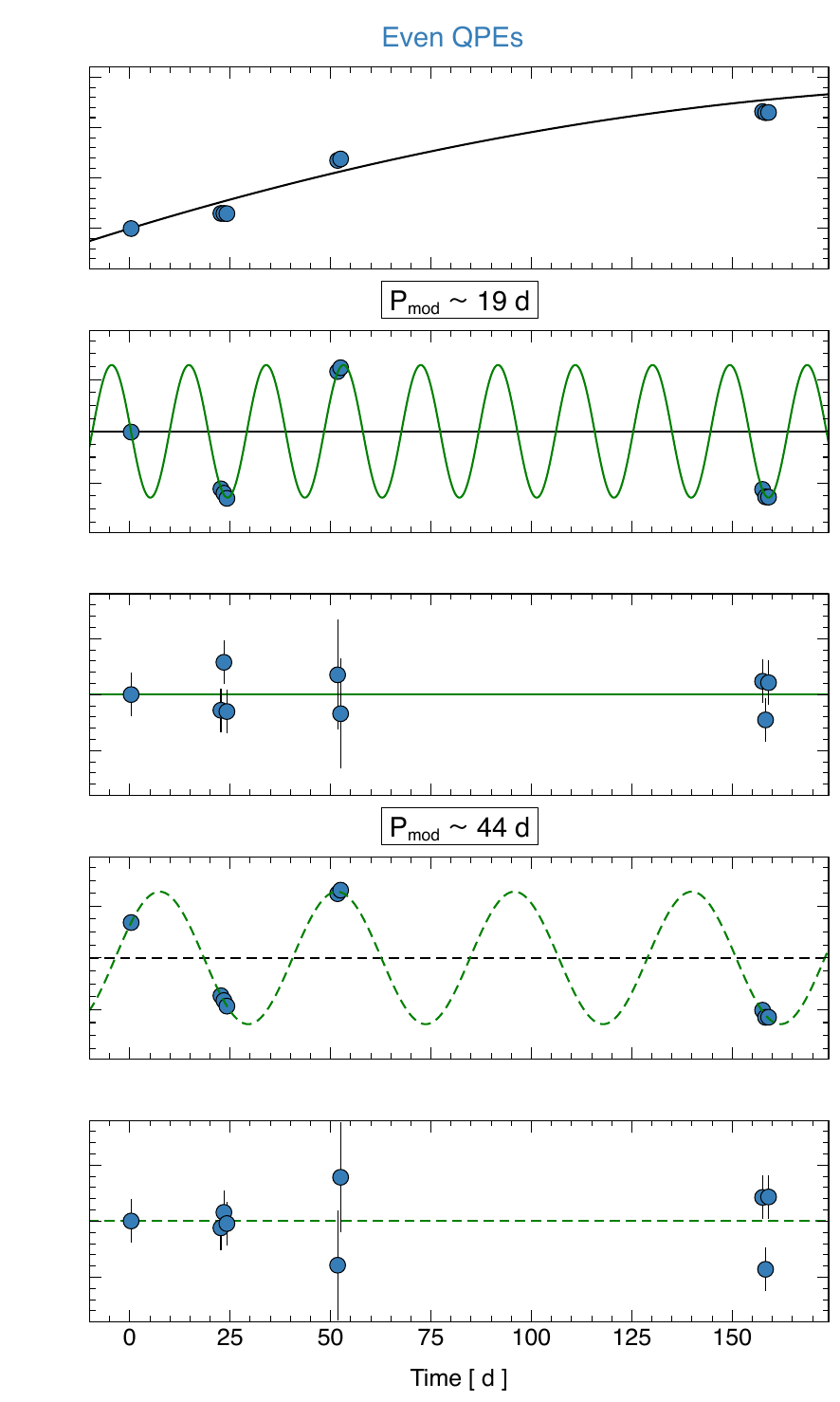}%
    \caption{O-C diagrams for GSN~069. We show the O-C diagrams for odd (left) and even (right) QPEs for GSN~069 resulting from identifying the first QPE of the May 2019 observation with the 211th even QPE. 
     The upper panels show the O-C data together with the linear plus parabolic baseline model for $P_{\rm mod}\simeq 19$~d. The lower panels show the corresponding residuals (O-C$_{\rm BASELINE}$) as well as the ones corresponding to the full best-fitting model including a sinusoidal modulation (O-C$_{\rm FULL}$) for the two possible $P_{\rm mod}$. The sinusoidal modulation is also shown in the O-C$_{\rm BASELINE}$ to guide the eye.}
    \label{fig:GSNOCmix}
\end{figure*}

A further quantity that is useful when dealing with periodic or quasi-periodic time series is the difference between Observed and Calculated (O-C) times of arrival of events (here QPEs) as a function of time (or epoch). ``O'' stands for the observed time of arrival of QPEs with respect to a chosen reference, while ``C'' is their expected time of arrival assuming that events are all separated by the same trial period $P_{\rm trial}$. The O-C analysis is a powerful tool to identify deviations from strictly periodic behaviour and is widely applied in the analysis of time series from variable stars, eclipsing binaries, or exoplanet's transits \citep[for basic definitions, see][]{2005ASPC..335....3S}. A strictly periodic time series produces linear O-C diagrams, where the slope is related to the difference between the true period $P_{\rm true}$ and the trial one $P_{\rm trial}$. The presence of a constant period derivative $\dot{P}$ produces an additional parabolic term in O-C diagrams, and its coefficient can be used to estimate the actual $\dot P_{\rm true}$. Besides these standard functional forms, O-C diagrams can help identify further deviations from periodic behaviour. Within the impacts model scenario, O-C diagrams need to be computed for odd and even QPEs separately as those are the events that are  expected to be separated by a constant period (the EMRI orbital one) at least at the Keplerian level while, as already mentioned, consecutive QPEs of different parity are generally separated by alternating longer and shorter intervals.

\subsection{O-C diagrams for GSN~069}
\label{sec:gsnOC}

Before discussing O-C diagrams for GSN~069, some caveats must be spelled out. In order to construct  O-C diagrams, one has to first select a reference event together with an assumed trial period $P_{\rm trial}$, and then derive the correct epoch (i.e. number of elapsed $P_{\rm trial}$) for all other events with respect to the reference one. For continuous time series this is a trivial exercise, but the GSN~069 data considered here comprise four observations with typical duration of the order of $0.5-1.5$~d separated by $\sim 23$~d, $\sim 29$~d, and $\sim 107$~d respectively. The data gaps introduce some degree of ambiguity in the correct QPE identification (number of elapsed $P_{\rm trial}$ with respect to the reference one).

We have therefore constructed a number of different versions of O-C diagrams for different QPE identifications, We defined the first two QPEs observed in December 2018 (see Fig.~\ref{fig:gsn}) as reference for odd and even QPEs respectively, and we assumed as $P_{\rm trial}$ the average measured $P_{\rm app} = 17.88$~hr throughout the $\simeq 160$~d campaign. QPEs belonging to the January 2019 observation are unambiguously identified, while those detected in February and May 2019 cannot be uniquely associated with the odd or even time series and with a specific number of elapsed $P_{\rm trial}$. The uncertainties in QPE epoch and parity are unavoidable and due to the sparse nature of the data and the relatively long gaps with respect to $P_{\rm trial}$. 

However, in this work we are interested in comparing the QPE timing properties with the specific impacts model which implies that odd and even QPEs need to share the same period (and period derivative, if present) as this is uniquely associated with the EMRI orbital period $P_{\rm orb}$ (and $\dot P_{\rm orb}$). Hence, at zeroth-order, the O-C data for odd and even QPEs must be described by standard O-C functional forms (linear or linear plus parabolic) with the same coefficients. We therefore consider as acceptable only O-C diagrams that fulfil this condition. By imposing this requirement, the ambiguity in the identification of QPEs was significantly reduced, at the expense that results presented below are not entirely model-independent. 

Despite our assumptions, the identification of QPEs from the last, May 2019 observation, the one associated with the longest gap in the data, remains uncertain since different identifications produce O-C diagrams that comply with our requirements. This ambiguity is associated with two possible sources of error: if an event (from the May 2019 data set) is associated with the odd time series but actually belongs to the even one (or vice-versa), all O-C data for that observation are shifted upwards or downwards by the average time separation between consecutive odd and even QPEs ($T_{\rm rec}$), that is half $P_{\rm trial}$. On the other hand, if the parity assigned to a given event is correct but the epoch (number of elapsed $P_{\rm trial}$) misidentified by one, all data points shift by one $P_{\rm trial}$. We have therefore produced three versions of the O-C diagrams for GSN~069 that differ by the identification of QPEs during the May 2019 observation. We discuss here the identification that produces intermediate O-C values for the May 2019 observation, while two other possibilities, for which O-C data are shifted upwards or downwards with respect to the case discussed here, are presented in Appendix~\ref{sec:GSNOCdiffID}. 

The upper panels of Fig.~\ref{fig:GSNOCmix} show the resulting O-C data for odd (left) and even (right) QPEs, where events of different parity are shown in separate panels for visual clarity (data points overlap at the chosen symbol size). We have then applied standard O-C models to the data: as mentioned, a linear trend is expected for strictly periodic time series, while the addition of a parabolic decay signals the likely presence of a period derivative. It was immediately clear that the data could be described by a linear plus parabolic trend at zeroth-order, but that a sinusoidal-like modulation was also likely present. A model of the form $a+bx+cx^2+A_{\rm mod}\sin(P_{\rm mod},\phi_{\rm mod})$ was found to describe the data well, although the sparse nature of the data prevented us to distinguish between two possible modulating periods of $\sim 19$~d and $\sim 43$-$44$~d. It is worth noting that, due to the limited number of data points in the odd QPEs time series (six, as are the free parameters of the adopted model), the model could not be formally applied in that case. The adopted fitting procedure is described in Section~\ref{sec:fitprocedure}.

\begin{figure}[t]
\centering \includegraphics[width=0.98\columnwidth]{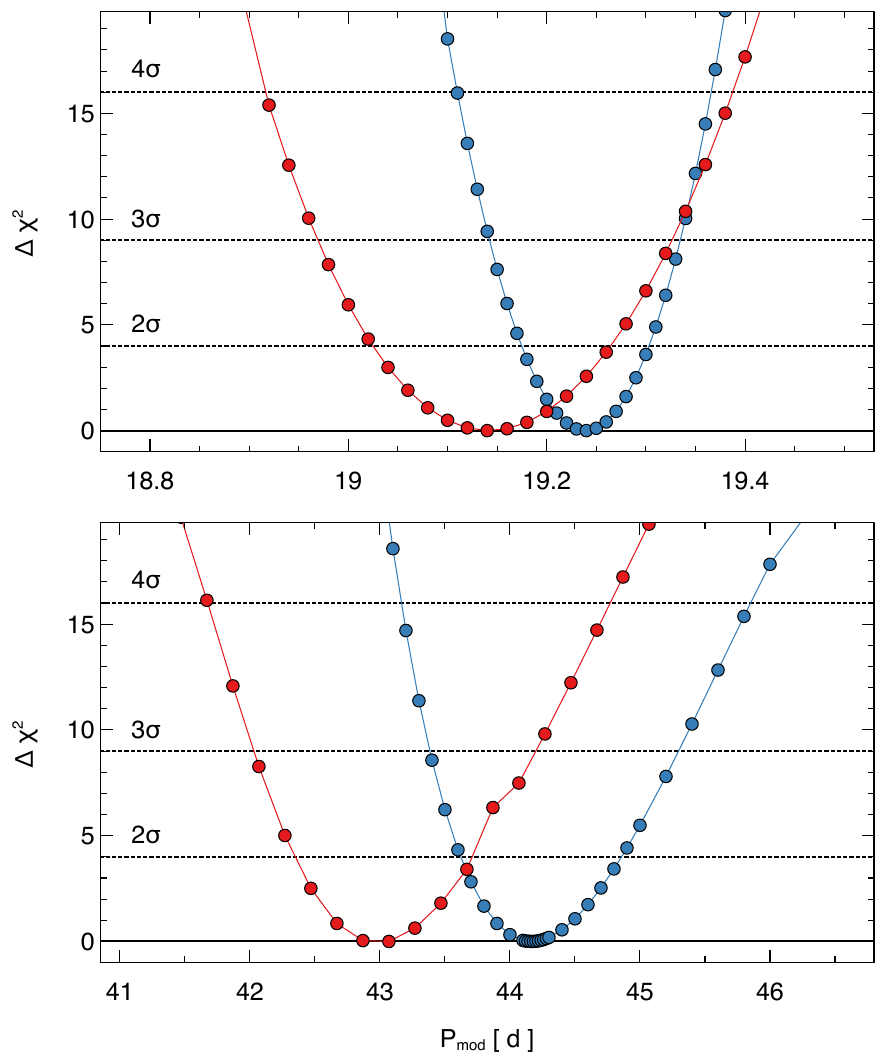}
%{\vspace{-0.2cm}}
\caption{$P_{\rm mod}$ detection in the O-C diagrams. We show $\Delta\chi^2$ as a function of modulating period $P_{\rm mod}$ for the final best-fitting model to the O-C diagrams in Fig.~\ref{fig:GSNOCmix}. Odd QPEs are shown in red, even ones in blue. We explored a wide range of $P_{\rm mod}$ between $4$~d and $90$~d, and all other local minima are noisy (i.e. neighbouring data points to the local minimum have much higher $\Delta\chi^2$ value).}
\label{fig:Pchi}
\end{figure}

The solid line in the upper panels of Fig.~\ref{fig:GSNOCmix} is the linear plus parabolic part of the best-fitting relation for the $P_{\rm mod}\sim 19$~d modulation (the relation for the $\sim 43$-$44$~d modulation is consistent with it within errors and is not shown for visual clarity). In the lower panels, we show the O-C residuals once the baseline linear plus parabolic best-fitting model is subtracted (O-C$_{\rm BASELINE}$) as well as those resulting from the subtraction of the full best-fitting model (O-C$_{\rm FULL}$). For both possible $P_{\rm mod}$, the shape of O-C$_{\rm BASELINE}$ is well described by a sine function (shown to guide the eye), as demonstrated by the small residuals shown in O-C$_{\rm FULL}$ in both cases. Results from the O-C analysis are given in Table~\ref{tab:OCgsnmix}, where we report as free parameters the physical $P_{\rm orb}$ and $\dot{P}_{\rm orb}$ (the EMRI orbital period and derivative) that are derived unambiguously from the coefficients of the linear and quadratic terms of the adopted model. The two best-fitting modulating periods are well defined in $\Delta\chi^2$ space, as shown in Fig.~\ref{fig:Pchi}. Other local minima in $\Delta\chi^2$ space were not well behaved, with large fluctuations around the local minimum.

\begin{table}[t]
        \centering
        \caption{ O-C analysis for GSN~069.}
        \label{tab:OCgsnmix}
        \begin{tabular}{lcc} % four columns, alignment for each                                      
          \hline
\T  & Odd &  Even  \B \\
\hline          
\T $P_{\rm mod} \sim 19$~d & &   \B \\
\hline
\T $P_{\rm orb}$~[hr] & $18.07\pm 0.05$ & $18.06\pm 0.03$ \B \\
\T $\dot{P}_{\rm orb}$~[$10^{-5}$] & $-3.6\pm 0.9$ & $-3.2\pm 0.6$ \B \\
\T $P_{\rm mod}$~[d] & $19.1\pm 0.1$ & $19.24\pm 0.06$ \B \\
\T A$_{\rm mod}$~[hr] & $2.60\pm 0.13$ & $2.57\pm 0.08$ \B\\
\hline
\T $P_{\rm mod} \sim 43$-$44$~d & &  \B \\
\hline
\T $P_{\rm orb}$~[hr] & $18.06\pm 0.06$ & $18.08\pm 0.04$ \B \\
\T $\dot{P}_{\rm orb}$~[$10^{-5}$] & $-3.0\pm 1.2$ & $-4.1\pm 0.8$ \B \\
\T $P_{\rm mod}$~[d] & $43.0\pm 0.7$ & $44.2\pm 0.6$ \B \\
\T A$_{\rm mod}$~[hr] & $2.8\pm 0.5$ & $2.7\pm 0.3$ \B \\
\hline
        \end{tabular}
        \tablefoot{$P_{\rm orb}$ is in fact the the draconitic period which, however, can be taken as an estimate of the orbital period to within few per cent, see text for details. $P_{\rm mod}$ and $A_{\rm mod}$ are the period and semi-amplitude of the O-C modulation, described here as a sine function. We report $2\sigma$ uncertainties on the best-fitting parameters stressing, once again, that the actual uncertainties might be larger due to systematic errors. The corresponding O-C diagrams are shown in Fig.~\ref{fig:GSNOCmix}. }  
\end{table}

The $\sim 19$~d modulating periods for odd and even QPEs are consistent with each other at the $2\sigma$ level, while only marginally so for the longer period. However, as already mentioned in Section~\ref{sec:definitions}, we stress again that the uncertainties reported in Table~\ref{tab:OCgsnmix} (and shown in Fig.~\ref{fig:Pchi} for the case of $P_{\rm mod}$) are statistical only. Within the framework of the impacts model, the O-C analysis should be carried over the actual (unknown) impacts times, while we have used the QPE (peak) times of arrival. This introduces systematic errors, as there is no guarantee that the delay between an impact and the peak of the corresponding X-ray emission is impact-independent. Hence, we caution that the uncertainties reported in Table~\ref{tab:OCgsnmix} are likely underestimated, and we discourage to consider our measurements as accurate at better than the few per cent level. This is the reason why we accept as plausible the $43$-$44$~d modulation despite a marginal discrepancy in the period derived from odd and even QPEs time series.

For both odd and even QPEs, the coefficients of the linear and quadratic terms are consistent with each other regardless of the actual modulating period, indicating that odd and even QPEs have common period $P$ and period derivative $\dot{P}$. The existence of such a solution is fully consistent with, and actually provides some support to, the impacts model because the QPE timing properties of both odd and even QPEs are imposed by the EMRI orbital period and its evolution. We point out that, within the impacts model scenario, the period derived from the O-C analysis is that of consecutive impacts at the same node (draconitic period), rather than the EMRI orbital period. In General Relativity, the two periods do not coincide as is the case for Keplerian orbits. However, the difference (due to relativistic corrections associated primarily with apsidal precession) is typically of the order of only few per cent for orbits wider than few tens $R_g$ as is almost certainly the case here \citep{2023ApJ...957...34L,2023A&A...675A.100F}, so that the derived period and period derivative can be considered as estimates of the EMRI $P_{\rm orb}$ and $\dot{P}_{\rm orb}$ to within few per cent. 

As reported in Table~\ref{tab:OCgsnmix}, we measure an average $P_{\rm orb} \simeq 18.07$~hr and $\dot{P}_{\rm orb} \simeq -3$-$4\times 10^{-5}$ in GSN~069. The O-C data are modulated on either a $\simeq 19$~d or $\simeq 43$-$44$~d timescale with semi-amplitude of the order of $2.5$-$2.8$~hr. As discussed in Section~\ref{sec:GSNOCdiffID}, the other two possible identifications for the May 2019 QPEs lead to consistent results for all parameters except $\dot{P}_{\rm orb}$ which takes the values of $\dot{P}_{\rm orb} = 0$ or $\dot{P}_{\rm orb} \simeq -6$-$7\times 10^{-5}$ depending on QPE identification. In summary, the EMRI orbital period and the properties of the O-C modulation (although with two possible solutions) are found to be robust against different QPE identifications. The only significant impact is on the derived $\dot{P}_{\rm orb}$ for which we suggest three possible values, depending on QPE identification in the May 2019 data, namely $\dot{P}_{\rm orb} = 0$, $-3$-$4\times 10^{-5}$, or $-6$-$7\times 10^{-5}$.

On the other hand, while parabolic trends in O-C diagrams are ubiquitously interpreted as a period derivative, we point out that this interpretation is not necessarily unique. If the O-C data were in fact modulated also on a further much longer timescale $P_{\rm mod}^{\rm long}$, their analysis on a baseline significantly shorter than $P_{\rm mod}^{\rm long}$ could result in the detection of spurious parabolic trends. As an example, if the central SMBH was spinning, nodal precession of the EMRI orbit would modulate the O-C diagrams on a very long timescale, significantly longer than the $\sim 160$~d baseline of the GSN~069 data (see Fig.~\ref{fig:spin} and associated discussion). We therefore caution that the $\dot{P}_{\rm orb}$ values derived above assume that the parabolic trend seen in the O-C diagrams is real and not just the sign of a longer timescale modulation.

\subsection{The 2023 campaign: hints for period decay}
\label{sec:Pdotconfirm}

QPE data obtained at significantly later times than 2019 could in principle be used to confirm (or reject) a period decay in GSN~069, perhaps also clarifying the correct version of the three O-C diagrams we have presented above and in Appendix~\ref{sec:GSNOCdiffID}. As mentioned, after May 2019, GSN~069 experienced a significant X-ray rebrightening and subsequent decay. QPEs disappeared at peak and had irregular properties during rise and decay. These QPEs are not very useful for deriving an averaged period to compare with the 2018-2019 one precisely because of their irregular timing behaviour which is likely due to changes in the disc structure and dynamics at rebrightening (possibly a second partial TDE) rather than in the EMRI's orbital parameters \citep[see also in][]{2024arXiv240412421L}.

\begin{figure}[t]
\centering \includegraphics[width=0.9\columnwidth]{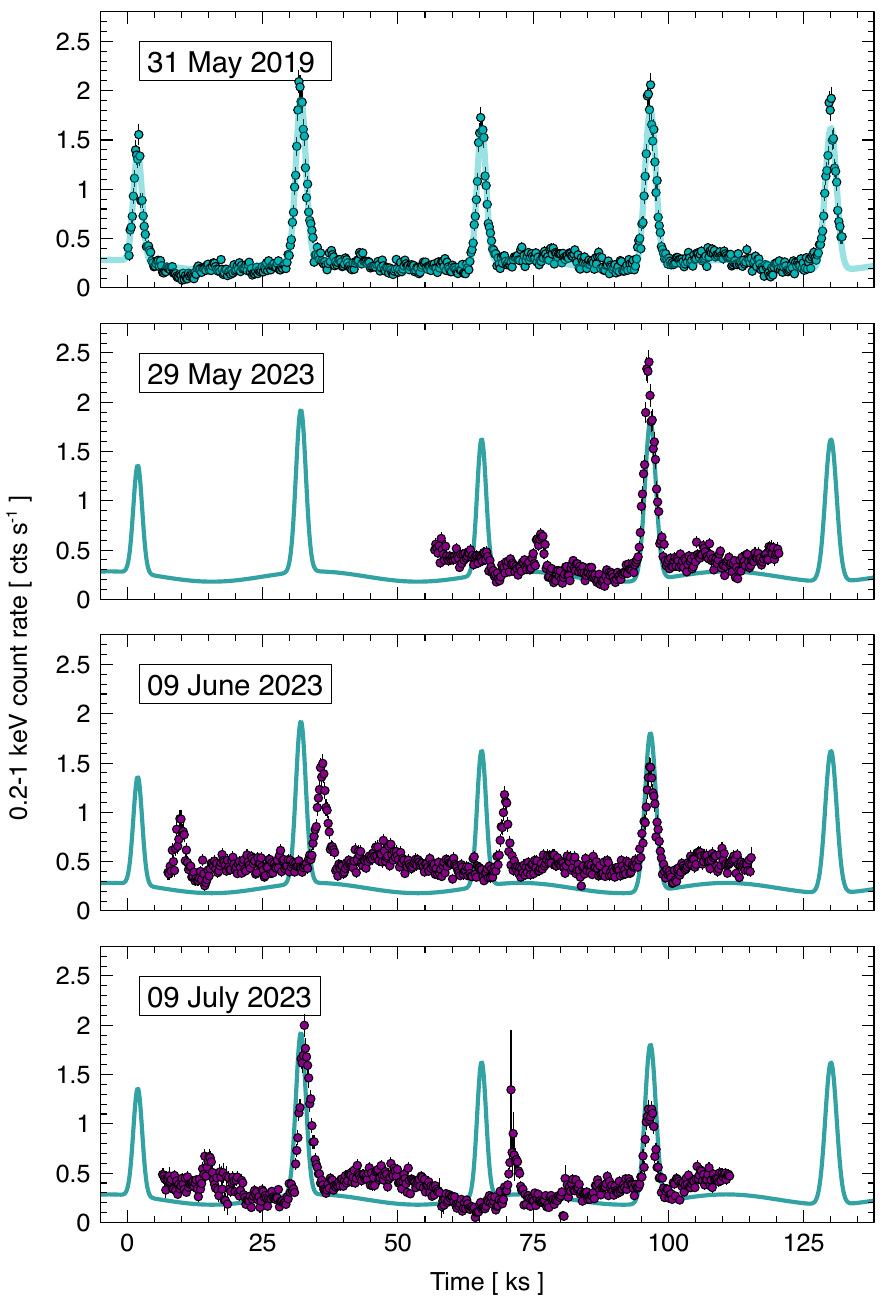}
\caption{The 2023 campaign on GSN~069. In the upper panel, we show the \xmm\ EPIC-pn light curve from the May 2019 observation, together with a representative best-fitting model. The EPIC-pn light curves from the 2023 campaign are shown in the lower panels. We aligned the last QPEs of these latter observations with one of the QPEs in the upper panel to ease comparison, and we also reproduced the best-fitting model for the May 2019 light curve in the lower panels. }
\label{fig:Pdot}
\end{figure}

GSN~069 was later re-observed three times by \xmm\ within $\sim 1.5$~months in 2023, about $3.5$~yr after rebrightening. QPEs were detected in all observations and appeared to be more regular than during rebrightening. The \xmm\ EPIC-pn light curves from these three observations are shown in Fig.~\ref{fig:Pdot}, where they are also compared with the May 2019 light curve (upper panel). Although still somewhat irregular, the timing pattern appears to approach that of the previous regular phase, represented here by the May 2019 light curve. The alternating $T_{\rm rec}$ behaviour is preserved, and longer recurrence times always follow stronger QPEs, as was the case in the regular phase (see Fig.~\ref{fig:gsn}). As visually clear from Fig.~\ref{fig:Pdot}, the time interval between QPEs of the same parity ($P_{\rm app}$) during the 2023 observations is generally shorter than during the May 2019 observation, which could indicate a period decay between the 2019 and 2023 epochs. 

The number of QPEs detected during the 2023 campaign is insufficient to perform a reliable O-C analysis. However, the light curves in Fig.~\ref{fig:Pdot} can be used to derive the average $P_{\rm app}$ in 2023. As shown in the previous O-C analysis (and confirmed in Appendix~\ref{sec:GSNOCdiffID}), the difference between $<P_{\rm app}> = P_{\rm trial}$ and $P_{\rm orb}$ was found to be of the order of $1$\% only in 2019. We assume here that the average $\left< P_{\rm app}^{\rm (2023)} \right> \simeq 16.7$~hr can be taken as representative of $P_{\rm orb}^{\rm (2023)}$ but with a larger uncertainty of $3$\% in the attempt to account for the limited number of independent measurements (4) and the somewhat still irregular nature of the 2023 QPE timing. We then derive $P_{\rm orb}^{\rm (2023)} = \left< P_{\rm app} \right>^{\rm (2023)}(1\pm 0.03)= 16.7\pm 0.5$~hr, so that the difference between the estimated orbital periods in 2023 and 2018-2019 is $\Delta P_{\rm orb} = -1.4 \pm 0.5$~hr. By taking as reference the mid-point of the time spanned by the 2018-2019 and 2023 observations, the elapsed time between the two  campaigns is $\Delta T \simeq 1560$~d, so that we estimate $\Delta P_{\rm orb}/ \Delta T = (-3.7 \pm 1.3 )\times 10^{-5}$, or $-0.3\pm 0.1$~hr per year. This is consistent with the inferred $\dot{P}_{\rm orb}^{\rm (2019)} = -3$-$4\times 10^{-5}$ from the O-C analysis presented above, and provides some support to results in Table~\ref{tab:OCgsnmix} with respect to alternative O-C identifications (see Appendix~\ref{sec:GSNOCdiffID}). Future monitoring campaigns detecting a sufficiently large number of QPEs to perform detailed O-C analysis on relatively long baselines will be extremely valuable to confirm (or reject) the suggested EMRI period derivative in GSN~069 that we consider, at present, tentative only.

\section{The impacts model: timing properties and comparison with GSN~069 data}
\label{sec:model}

In order to show the general properties of the impacts model, and to compare its predictions with the GSN~069 data, we considered simulations performed using the code developed by \citet{2023A&A...675A.100F} to which we refer for details (see also Appendix~\ref{sec:notesmodel}). Although disc precession can be implemented in their code, we initially switched it off to illustrate the general behaviour of the QPE timing within the context of the simplest version of the impacts model. 

We selected fiducial parameters for GSN~069 assuming a non-spinning  primary SMBH with mass $M = 8\times 10^5~$M$_\odot$ forming an EMRI system with a secondary orbiting object of mass $m \ll M$. The EMRI's orbital period was set to $18$~hr, and the orbital eccentricity to $e=0.05$, of the order of those inferred in GSN~069 within the impacts model scenario \citep{2023A&A...675A.100F,2024PhRvD.110h3019Z}. Choosing a different black hole mass keeping an orbital period of $18$~hr (as set by the data) only changes the EMRI semi-major axis and, as a consequence, the apsidal precession timescale without affecting in any way the general behaviour presented below. 

We assumed that the accretion disc around the primary SMBH has angular momentum misaligned by $i_{\rm disc} = 5^\circ$ with respect to the z-axis\footnote{In absence of disc or nodal precession, the only relevant angle is that between the disc and the EMRI's orbit angular momenta, so having a disc inclined with respect to the equatorial plane is not necessary here. However, the same set-up will be used in Section~\ref{sec:discprecession} when discussing the effects of disc precession, so that we introduce the adopted geometry here.}, while the EMRI angular momentum vector was set at $i_{\rm EMRI} = 10^\circ$. The observer inclination with respect to the z-axis was set to $i_{\rm obs} = 30^\circ$. We point out that different choices of $i_{\rm obs}$ do not significantly affect the results discussed below. As mentioned, no disc precession was included initially and we assumed, for simplicity, $\dot{P}=0$.

\begin{figure}[t] 
    \centering
        \includegraphics[width=0.45\textwidth]{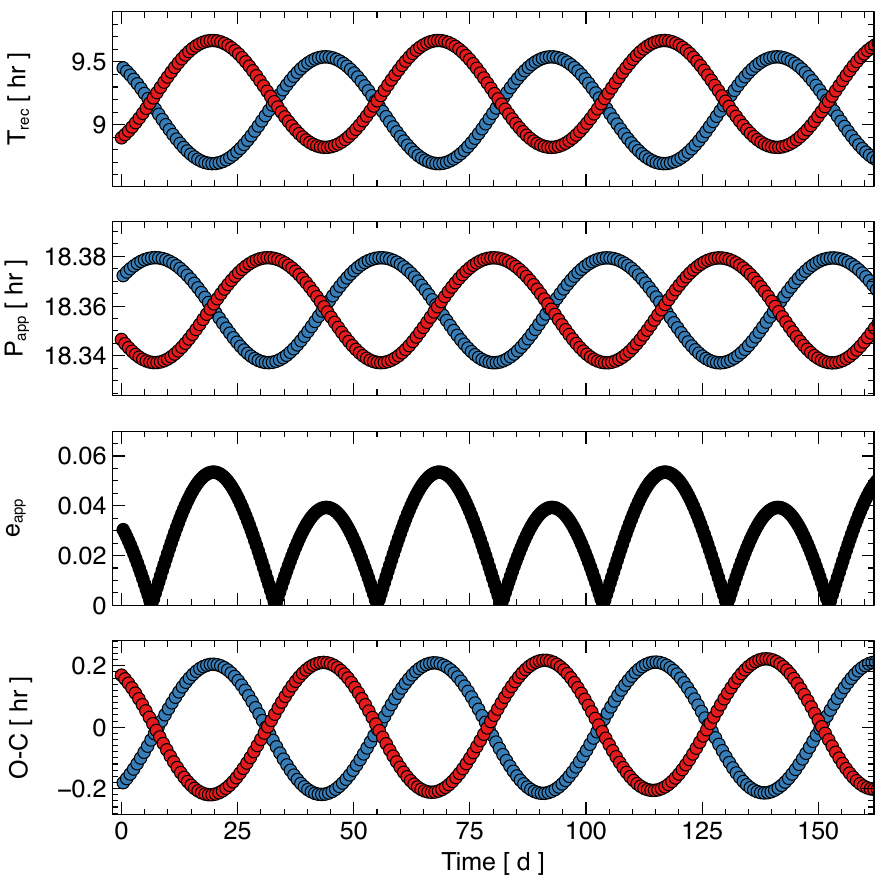}%
    \caption{QPE timing from the impacts model. We show $T_{\rm rec}$, $P_{\rm app}$, $e_{\rm app}$, and the the O-C diagrams for a nearly circular EMRI orbit with eccentricity $e=0.05$, and parameters commensurate with those of GSN~069. Odd and even QPEs are shown in red and blue respectively. We point out that the data points in the lower panels (O-C diagrams) are not exactly aligned with those in the upper ones. This is because the abscissa of a given data point in O-C diagrams is a multiple of $P_{\rm trial}$ rather than the observed QPE arrival time.}
    \label{fig:ex}
\end{figure}

In Fig.~\ref{fig:ex}, we show $T_{\rm rec}$, $P_{\rm app}$, $e_{\rm app}$ (defined in Eq.~\ref{eq:T_rec}-\ref{eq:e_app}) and the O-C diagrams from the simulated light curves\footnote{As discussed in Section~\ref{sec:gsnOC}, the apparent period $P_{\rm app}$ is actually the draconitic period rather than the orbital one. As a consequence, the average $P_{\rm app}$ in Fig.~\ref{fig:ex} is not equal to the Keplerian orbital period of $18$~hr, but longer by about $2$\%.}. Odd and even QPEs are represented with different colours. $T_{\rm rec}$ and $P_{\rm app}$ for the two branches are in phase opposition and periodic at the apsidal precession timescale, as discussed e.g. by \citet{2024MNRAS.527.4317L}. The anti-correlation of $P_{\rm app}$ propagates into that of the O-C diagrams for the two branches shown in the lower panels of Fig.~\ref{fig:ex}. This is a well known result in the analysis of O-C diagrams of eclipsing binaries showing apsidal motion, where the anti-correlation between the O-C diagrams of primary and secondary eclipses is ubiquitously observed \citep{2014A&A...572A..71Z}. Due to the symmetric nature of its definition, the distinction between odd and even QPEs disappears in $e_{\rm app}$ which exhibits a distinctive bell-like shape spanning all allowed values from zero to $\sim$$e_{\rm app}^{\rm max}$ and reaching $e_{\rm app}=0$ twice per apsidal period (at each crossing between the  $T_{\rm rec}$, see upper panels of Fig.~\ref{fig:ex}) as illustrated in Fig.~\ref{fig:schema}.

The properties outlined above are not affected significantly by changing any of the system parameters. The only qualitative effect is that induced by a non-zero primary SMBH spin because, even without considering any disc precession, it introduces nodal precession of the EMRI orbital plane, thus affecting the QPE timing. The effects of black hole spin are briefly discussed in Appendix~\ref{sec:notesmodel}, and they are ignored in our work because the nodal precession timescale is, for any reasonable choice of parameters, much longer than the apsidal precession one and than that spanned by typical QPE observations, and is of the order of $\simeq 1$~yr \citep{2023ApJ...957...34L}. Some minor effects are also present on shorter timescales, but they do not modify the qualitative behaviour of the QPE timing we are interested in here, at least on timescales probed by observations. We therefore selected a non-spinning primary SMBH throughout our work, but stress that future implementations attempting to derive physical parameters from the data should also include black hole spin effects self-consistently.

The model's predictions shown in Fig.~\ref{fig:ex} can be compared with the observed data shown in Fig.~\ref{fig:gsn} ($T_{\rm rec}$, $P_{\rm app}$, $e_{\rm app}$) and with the O-C diagrams of Fig.~\ref{fig:GSNOCmix} (see also Fig.~\ref{fig:GSNOClin} and \ref{fig:GSNOCquad}). The most striking discrepancy between the model's prediction and QPE data is that the observed apparent orbital period $P_{\rm app}$ as well as the O-C diagrams for odd and even QPEs are not in phase opposition. The number of observed consecutive $P_{\rm app}$ data points is limited (see Fig.~\ref{fig:gsn}), but for the data to be consistent with the behaviour shown in Fig.~\ref{fig:ex}, we should have observed GSN~069 precisely when $P_{\rm app}$ of odd and even QPEs cross in January and May 2019, which appears highly unlikely. As per the O-C data, all three versions of the O-C diagrams that we consider acceptable are indeed modulated at a super-orbital timescale, but the two branches only have marginal phase difference, as clear from e.g. Fig.~\ref{fig:GSNOCmix}. Moreover, the observed amplitude of the O-C modulation ($\simeq 2.5$-$2.8$~hr) is about one order of magnitude higher than that induced by apsidal precession which is of the order of few minutes only (see Fig.~\ref{fig:ex}, and compare it with Fig.~\ref{fig:GSNOCmix} as well as with with Fig.~\ref{fig:GSNOClin} and \ref{fig:GSNOCquad}). We point out that none of the O-C diagrams that we have produced, including those that were eventually rejected for not fulfilling our requirements, exhibited odd and even QPEs O-C data in even approximate phase opposition. Further discrepancies are also present for the observed $T_{\rm rec}$ as they indeed alternate as expected from the model, but are not exactly anti-correlated or in phase opposition. As an example, the increase in $T_{\rm rec}$ for the lower branch in May 2019 (red data points in Fig.~\ref{fig:gsn}) is much more pronounced than its decay for the upper one (blue data points). Finally, the observed $e_{\rm app}$ does not not display the typical bell-like shape expected from the model, although this might be a consequence of the limited number of data points and of the relatively short timescales that could be explored continuously. 

On the other hand, some of the model expectations are indeed fulfilled by the data. The observed $T_{\rm rec}$ consistently alternate between longer and shorter recurrence times, as natural in the impacts model. Moreover, there exist O-C diagrams for which the coefficient of the linear (and parabolic, when needed) part of the best-fitting relation is precisely the same for both odd and even QPEs, indicating that odd and even QPEs share the same period (and period derivative, when present), consistent with expectations. It is therefore plausible that the impacts model needs to be modified rather than rejected altogether.
 
Impacts models with no disc precession, similar to the one discussed above, have been successfully applied by \citet{2021ApJ...921L..32X} and \citet{2024PhRvD.109j3031Z,2024PhRvD.110h3019Z} to multi-epoch observations of GSN~069. Besides the specific parameters (black hole mass and EMRI orbital parameters) that basically affect timescales rather than the qualitative behaviour, their model is equivalent to the one discussed above, whose properties are shown in Fig.~\ref{fig:ex}. Due to the discrepancies highlighted above, the data do not seem to comply with the model's version implemented in these works (i.e. impacts on a geometrically thin disc with no disc precession). On the other hand, \citet{2023A&A...675A.100F} have qualitatively reproduced one single-epoch light curve of GSN~069 (as well as of other QPE sources) by including a spinning primary SMBH with a misaligned and rigidly precessing accretion disc. Due to the additional ingredient of disc precession, their model cannot be directly compared with Fig.~\ref{fig:ex} and its ability to reproduce the observed properties highlighted above is studied in more detail in Section~\ref{sec:discprecession}.

\section{Possible way(s) forward}
\label{sec:wayforward}

Despite the ambiguities discussed in Section~\ref{sec:OC}, O-C diagrams retain the largest number of useful data points (one per QPE)  with respect to all other quantities we have introduced ($T_{\rm rec}$, $P_{\rm app}$, and $e_{\rm app}$). They are therefore more informative in deriving trends and in comparing the observed QPE timing properties with model's predictions, at least at the qualitative level we are interested in here. As an example, the four light curves in Fig.~\ref{fig:gsn} provide 15 O-C data points, but only 11 $T_{\rm rec}$ and 7 $P_{\rm app}$ and $e_{\rm app}$. 

Focusing on O-C diagrams, the fact that odd and even QPEs O-C data are modulated periodically with minimal phase difference means that QPEs belonging to one branch are delayed (or anticipated) by the same time interval and with the same sign, as those belonging to the other with respect to an assumed perfect periodicity. These common delays oscillate on a long, super-orbital timescales of $\simeq 19$~d or $\simeq 43$-$44$~d in GSN~069. A natural super-orbital timescale is the EMRI apsidal precession, but apsidal motion inevitably makes the two branches oscillate in phase opposition with each other (see Fig.~\ref{fig:ex}) so that, at least in the framework of the impacts model, apsidal precession cannot be associated with the observed O-C periodic modulation.

The only plausible way of introducing a correlation between the two branches is that there is a further modulating timescale that dominates, over the observed baseline, with respect to apsidal precession. In particular, such modulating timescale should be shorter than the apsidal one\footnote{If the modulation had longer timescale than apsidal precession, it will produce a correlation on long timescales, but the anti-correlation on short ones would remain, exactly as is the case for the modulation induced by the orbital nodal precession shown in Fig.~\ref{fig:spin}.}, and have sufficiently high amplitude for the anti-correlation induced by apsidal precession to be overwhelmed or, at least, to be less dominant.

Within the impacts model scenario, there is a natural way of introducing an external periodic modulation. As mentioned, when the primary SMBH is spinning, the orbital plane of the EMRI periodically changes inclination with respect to the plane of the disc due to nodal precession. As discussed in Appendix~\ref{sec:notesmodel} and shown in Fig.~\ref{fig:spin}, this induces a coherent modulation of the O-C diagrams for odd and even QPEs. However, the timescale associated with nodal precession is much longer than the apsidal precession one, so that the latter dominates on short timescales and O-C diagrams for odd and even QPEs are still in phase opposition over the baseline that can be currently probed by observations (say a few apsidal precession timescales). On the other hand, if the SMBH is spinning and the accretion disc misaligned, disc precession can also be present. The effect of disc precession on O-C diagrams is likely similar to that induced by nodal precession of the orbit (see Fig.~\ref{fig:spin}), but the disc precession timescale can be significantly shorter than the nodal and apsidal ones depending on black hole mass, spin, and disc structure \citep{2016MNRAS.455.1946F}, so that it might dominate over apsidal precession and break the expected anti-correlation of O-C diagrams for odd and even QPEs. We explore this possibility in Section~\ref{sec:discprecession} below. 

On the other hand, QPE light curves are remarkably similar to eclipsing binary ones, only having bursts of X-ray emission rather than eclipses, which in fact motivated initial attempts to model QPEs in terms of self-lensing in an SMBH binary \citep{2021MNRAS.503.1703I} and inspired us to apply the O-C technique to QPEs. One possible source for the observed O-C modulation comes directly from the analogy between the two systems. Apsidal motion is often seen in O-C diagrams of eclipsing binaries, and is in fact identified precisely by the anti-correlation between primary and secondary eclipses in O-C data. However, the O-C diagrams for primary and secondary eclipses do sometimes correlate, which is often identified with light-travel-time effects arising due to the motion of the binary system around the centre of mass with a third orbiting star. In fact, correlated and periodic O-C diagrams have been used to infer the presence of triple systems in many instances \citep[see, for example ][]{2015AJ....149..197Z}. The same idea can in principle apply to QPE data and would imply the presence of an outer SMBH forming a binary with the EMRI plus disc (QPE-emitting) system. Such a hierarchical triple system, comprising an outer SMBH binary and an inner EMRI, is discussed in Section~\ref{sec:triples} below.

In any case, a qualitative solution to the QPE timing behaviour in GSN~069 must: (i) generally preserve the alternating $T_{\rm rec}$, although introducing some distortion that breaks the (unobserved) anti-correlation between odd and even QPEs recurrence times; (ii) align the $P_{\rm app}$ for the two branches on the same (or similar) functional form, again breaking the expected anti-correlation on the apsidal precession timescale; (iii) produce periodic O-C diagrams at some super-orbital timescale in which O-C data for odd and even QPEs have only marginal phase difference.

\section{On the black hole mass in GSN~069}
\label{sec:BHmasses}

In the context of the impacts model, the mass of the EMRI's primary SMBH plays a crucial role as it sets the EMRI semi-major axis (once an orbital period is known or estimated) and thus also the apsidal precession timescale as a function of orbital eccentricity. Before discussing the two possible scenarios of disc precession and of an hierarchical triple  in some detail, it is therefore worth clarifying the current status on black hole mass estimates in this galactic nucleus.

Optical spectroscopic observations have been used to derive the central stellar velocity dispersion in GSN~069 as $\sigma_* = 63\pm 4$~km~s$^{-1}$ \citep{2022A&A...659L...2W,2024ApJ...970L..23W}. By using the $M-\sigma$ relation as derived by \citet{2011ApJ...739...28X} for low-mass active galaxies, one can estimate the associated total nuclear mass as $\log M_{\rm tot} = 6.0\pm 0.5$, where we have assumed a conservative $0.5$~dex uncertainty associated with the scatter in the $M-\sigma$ data, rather than the (roughly twice as small) statistical error. Here we define the QPE-emitting system as an EMRI in which the central SMBH has mass $M_1$, and its low-mass companion has mass $m\ll M_1$. The mass derived from the $M-\sigma$ relation refers to the total nuclear mass, so that $M_{\rm tot} = M_1 + m \simeq M_1$. On the other hand, in presence of a second nuclear SMBH with mass $M_2$, that is if a hierarchical triple system is present, $M_{\rm tot} = M_2 +M_1 +m \simeq M_2 + M_1$.  

While $M_{\rm tot}$ can be estimated through the $M-\sigma$ relation, $M_1$ is associated with accretion disc emission (and QPEs), so that an estimate on $M_1$ can in principle be derived from continuum spectroscopy, and then compared to $M_{\rm tot}$. A clear indication that $M_1 < M_{\rm tot}$ would signal the likely presence of a nuclear SMBH binary. However, any X-ray-based estimate of $M_1$ is subject to considerable systematic uncertainties. This is primarily because only a very restricted portion of the full spectral energy distribution (SED) is observed in the X-rays (the high-energy tail of the thermal disc emission). Optical and UV photometric data are severely contaminated by stellar light, and appropriate subtraction is rather uncertain. A detailed study of the SED of GSN~069, as well as of other QPE galactic nuclei, is beyond the scope of this work and is deferred to future studies (see, for example, promising work by \citealt{2024arXiv240817296G} on the TDE ASASSN-14li). Moreover, the presence of  ionised absorption, that affects the X-ray spectrum of GSN~069 \citep{2024arXiv240617105K}, introduces further model-dependent uncertainties \citep{2023A&A...670A..93M}. Finally, the X-ray part of the SED is also highly sensitive to the specific adopted accretion disc model through, for example, the assumed disc truncation radius, black hole spin, inclination, or colour-correction factor. We have nevertheless attempted to derive X-ray-based estimates for $M_1$ in GSN~069 using the {\texttt optxagnf} and {\texttt agnsed} X-ray spectral models (used here switching off all Comptonisation components), that basically differ by the adopted colour-correction for the disc emission \citep{2012MNRAS.420.1848D,2018MNRAS.480.1247K}, but we could never obtain uncertainties lower than the order of magnitude level. If the disc is assumed to reach the innermost stable circular orbit, the most important contribution to the mass error budget comes from the black hole spin value and sign, with the lowest black hole masses reached for maximally spinning Kerr black holes with retrograde accretion (and the highest for prograde accretion). The typical range we derive  is $\sim 10^5$-${\rm few}\times 10^6~M_\odot$. As the range is consistent with, but even wider than, that from the $M-\sigma$ relation, we do not discuss these estimates further as they are not very informative.

\section{Disc precession}
\label{sec:discprecession}

Let us first consider a simple toy model that helps anticipating what the effects of disc precession on O-C diagrams might be. To aid visualisation, we assume that, when non-precessing, the disc lies in the $x$-$y$ plane, and that the EMRI orbital plane is orthogonal to it. When precession is present with period $P_{\rm disc}$, the disc forms an angle $\theta$ (with respect to the $x$-$y$ plane) that is roughly constant over one orbital period for any $P_{\rm orb}\ll P_{\rm disc}$. Impacts at the ascending and descending nodes occurring at radius $R_{\rm asc,\,desc}$ with orbital velocity $v_{\rm asc,\,desc}$ are then delayed (or anticipated) with respect to the case of no disc precession by $\Delta t_{\rm asc,\,desc} \simeq (R/v)_{\rm asc,\,desc} \cdot \theta$. In order to have similar O-C modulating amplitude, impacts at the ascending and descending nodes must therefore satisfy $(R / v)_{\rm asc} \approx (R/v)_{\rm desc}$. This is, by definition,  always approximately the case for nearly circular orbits. On the other hand, when the EMRI eccentricity is significantly different from zero, this condition is satisfied only in a limited range of apsidal phases, the range during which the two nodes roughly align with the EMRI orbit's latus rectum and the apparent eccentricity $e_{\rm app}$ is the highest since, in this case, impacts occur roughly at the same radius and with similar orbital velocity. Disc precession might therefore generally account for O-C modulations for EMRIs in nearly circular orbits which is very likely the case in GSN~069, but only during a fraction of the apsidal period whenever the eccentricity is significantly non-zero. 

The numerical implementation of the impacts model by \citet{2023A&A...675A.100F} naturally includes rigid disc precession resulting from a TDE-like accreting flow misaligned with respect to the equatorial plane of a spinning central SMBH. In their formulation, the disc precession timescale is dictated by the Lense-Thirring frequency weighted by the disc's angular momentum over its radial extent \citep{2016MNRAS.455.1946F}. However, in order to explore the parameter space without the complications induced by the EMRI orbital nodal precession (Fig.~\ref{fig:spin}), which is still much slower than the other relevant precession timescales at play and can therefore be neglected at first order, we imposed here rigid disc precession at an arbitrarily chosen frequency maintaining the spin of the central SMBH equal to zero.  

\begin{figure}[t] 
    \centering
        \includegraphics[width=0.9\columnwidth]{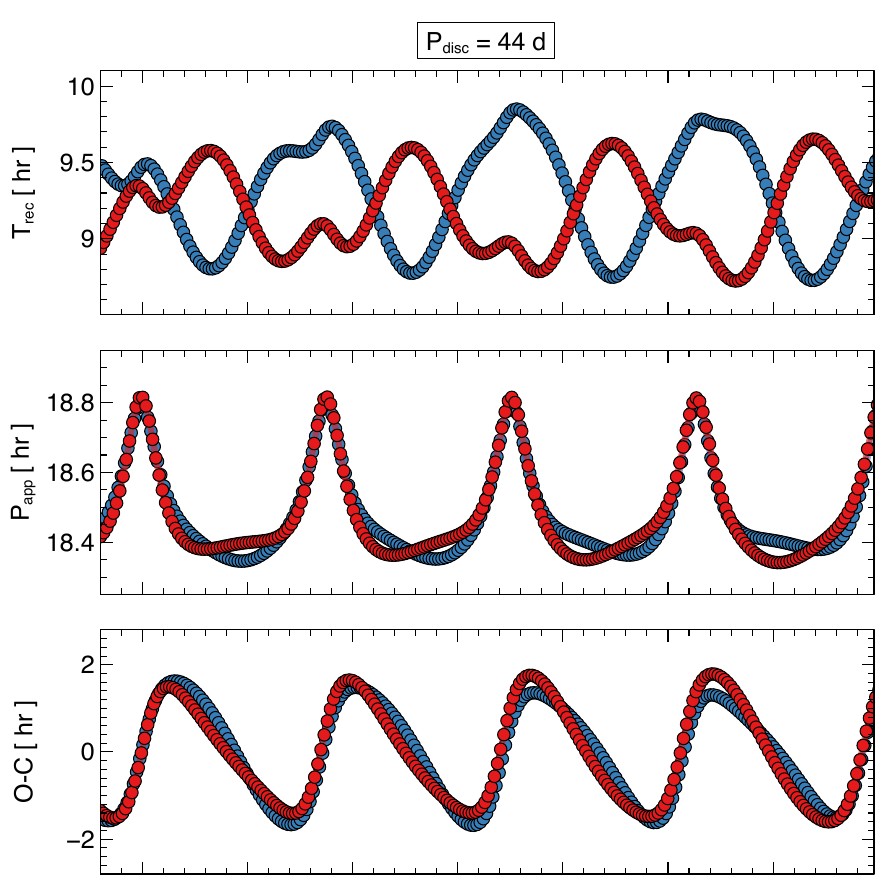}       
\includegraphics[width=0.9\columnwidth]{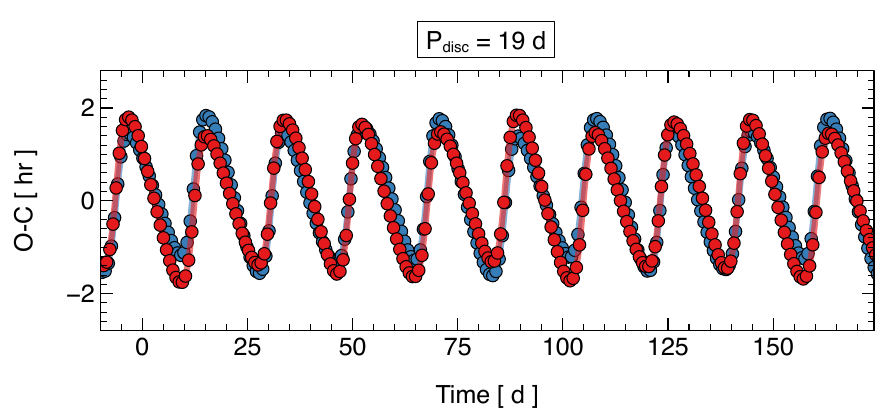}%
    \caption{Disc precession solution for GSN~069. In the upper three panels, we show $T_{\rm rec}$, $P_{\rm app}$, and the O-C diagrams for a disc precession solution with $P_{\rm disc}\sim 44$~d for GSN~069 (see text for further details). The lower panel shows the O-C diagrams for the same simulation but with $P_{\rm disc}\sim 19$~d.}
    \label{fig:GSNprecession}
\end{figure}

We selected parameters consistent with the observed QPE timing properties in GSN~069. In particular, we assumed a black hole mass of $8\times 10^5~M_\odot$, and an EMRI with $P_{\rm orb} \simeq 18$~hr and eccentricity $e=0.05$. The disc precession period was set to either $P_{\rm disc} =44$~d or $19$~d, representing the two possible modulation periods obtained from the O-C analysis. The other relevant parameters were set to the same values as those in Fig.~\ref{fig:ex}, namely $i_{\rm obs}=30^\circ$, $i_{\rm disc} = 5^\circ$, and $i_{\rm EMRI} = 10^\circ$ (all with respect to the $z$-axis), as we were interested in showing a qualitative match with the observed O-C data rather than in finding an accurate best-fit.

The resulting $T_{\rm rec}$, $P_{\rm app}$, and O-C diagrams are shown in the upper three panels of Fig.~\ref{fig:GSNprecession} for $P_{\rm disc} = 44$~d, while the lower panel only shows the O-C diagrams for $P_{\rm disc} = 19$~d. We chose to show $T_{\rm rec}$ and $P_{\rm app}$ for the case of the longer modulating timescale not because we believe it to be a more plausible solution, but rather for visual clarity, as the different quantities are less compressed on the $x$-axis. The variability of all quantities depends on the interplay between the disc and the apsidal precession timescales, but disc precession dominates, breaking the anti-correlation pattern that is present when only apsidal precession is considered (see Fig.~\ref{fig:ex}) and introducing instead correlated variability of $P_{\rm app}$ and O-C data. For both $P_{\rm disc}$, the simulated O-C diagrams shown in the two lower panels can be compared with the corresponding observed ones in Fig.~\ref{fig:GSNOCmix} (as well as in Fig.~\ref{fig:GSNOClin} and \ref{fig:GSNOCquad}). The two branches of the simulated O-C data are well correlated with minimal phase difference, and the modulating shape and period are in good agreement with the data for both the $44$~d and $19$~d disc precession periods. The modulation amplitude is somewhat under-estimated, but there is a good overall agreement. The modulation amplitude primarily depends on the disc misalignment $i_{\rm disc}$ with respect to the $z$-axis. Although we did not explore the full parameter space as we are here interested in the general behaviour rather than in finding accurate best-fitting solutions and parameters, we nevertheless report that a good match with the observed $\simeq 2.5$~hr modulation amplitude is reached by increasing $i_{\rm disc}$ from $5^\circ$ to $\simeq 20^\circ$. 

The simulated O-C diagrams are not perfectly sinusoidal, which is visually more evident in the $P_{\rm disc} = 44$~d case,  and this might actually be the origin of the ambiguity between the $43$~d or $\sim 44$~d modulation period in the odd and even QPEs data of GSN~069 since the latter was obtained by assuming a perfect sine function with sparse sampling. $T_{\rm rec}$ and $P_{\rm app}$ shown in the two upper panels of Fig.~\ref{fig:GSNprecession} also comply with the requirements that are necessary for the model to represent a qualitatively viable solution to the QPE timing: the alternating recurrence times are preserved, but the perfect anti-correlation that is expected from the impacts model with no disc precession is broken; the  $P_{\rm app}$ for odd and even QPEs approximately align on the same function, rather than being in phase opposition.

As mentioned, \citet{2023A&A...675A.100F} have included disc precession in their impacts model to reproduce single-epoch light curves of GSN~069, RX~J1301.9+2747, eRO-QPE1, and eRO-QPE2. As noted in their work, the disc precession is much longer than that of the typical single-epoch observation (about $1.5$~d), so that its effects on QPEs timing cannot be seen in their light curves. For GSN~069 they suggest a SMBH spin $\chi = 0.1$ which, combined with a black hole mass of $10^6~M_\odot$, an EMRI semi-major axis of $\simeq 160~R_g$ with orbital period $P_{\rm orb} \simeq 18$~hr, and the assumed disc properties leads to $P_{\rm disc} \simeq 125$~d. Their modelling represents the first attempt to derive a self-consistent solution for the QPEs timing (as well as X-ray peak luminosity) including an external modulation, and can be considered  successful at the qualitative level with the discrepancies (e.g. $P_{\rm disc} \simeq 125$~d instead or $\simeq 20$~d or $\simeq 43$-$44$~d) being likely only due to the limited baseline (single-epoch observations) used in their analysis.

We conclude that a rigidly precessing disc with precession period $P_{\rm disc} = 19$~d or $43$-$44$~d represents a viable mechanism by which the impacts model can be reconciled with the observed QPE timing data in GSN 069. It is also worth mentioning that, according to the study by \citet{2016MNRAS.455.1946F}, disc precession timescales of the order of $\sim 20$-$40$~d can be reached for dimensionless black hole spin of the order of $0.1$-$0.6$ for the relevant range of black hole masses. On the other hand, the disc precession timescale also depends on disc extent and structure (e.g. viscosity), so that deriving an estimate of the SMBH spin is not trivial.

\section{Hierarchical triple: an outer SMBH binary and inner EMRI system}
\label{sec:triples}

If the EMRI-system we have considered so far was a member of a (hierarchical) triple system comprising an outer SMBH binary, orbital motion of the QPE-emitting inner EMRI around the centre of mass with the second SMBH would induce time delays in the time of arrivals of QPEs. Odd and even QPEs would be modulated in roughly the same way, as QPEs delays are simply associated with the light travel time from the impact to the observer, which is a function of the outer binary orbital phase and observer inclination. Within this scenario, $P_{\rm orb}$ and $P_{\rm mod}$ in Table~\ref{tab:OCgsnmix} are estimates of the inner, QPE-emitting EMRI orbital period, and the outer binary one respectively ($P_{\rm out}$), while the amplitude of the O-C modulation (together with the observer inclination) sets the geometrical scale of the outer binary. 

Since the O-C modulation in GSN~069 is consistent with a sine function, we assume for simplicity that the outer binary is on a circular orbit, although the sparse nature of the O-C data might allow for more complex functional forms associated with an eccentric outer binary. The orbital radius of the EMRI (or, equivalently of the SMBH with mass $M_1$) around the centre of mass with $M_2$ is then $a_1 = A_{\rm mod}c/\sin i_{\rm obs}$, where $i_{\rm obs}$ is the angle between the observer line of sight and the outer binary angular momentum. $a_1$  is related to the binary separation $a_{\rm out}$ by $a_{\rm out}=a_1~(1+q)$, where $q=M_1/M_2 \leq 1$ and $M_2$ is the outer SMBH mass. By using Kepler's third law, one can then derive the total mass $M_{\rm tot} = M_1 +M_2 +m \simeq M_1 + M_2$ as a function of $P_{\rm mod}$, $A_{\rm mod}$, $i_{\rm obs}$, and $q$ as

\begin{align} 
M_{\rm tot} & = \frac{4\pi^2}{G} \frac{a_{\rm out}^3}{P_{\rm out}^2}
 = \frac{4\pi^2 c^3}{G} \left( \frac{1+q}{\sin i_{\rm obs}}\right)^3 \frac{A_{\rm mod}^3}{P_{\rm mod}^2}~. \label{eq:Kep}
\end{align}
Eq.~\ref{eq:Kep} can then be used as a consistency check for the hierarchical triple hypothesis in GSN~069 by requiring that the $M_{\rm tot}$ derived by considering the observed upper limit on $P_{\rm mod}$ and lower limit on $A_{\rm mod}$ does not exceed the upper limit from the $M-\sigma$ relation.  As mentioned, the uncertainties reported in Table~\ref{tab:OCgsnmix} are statistical only, and our measurements are likely to be subject to some systematic uncertainty due to the unknown, and possibly impact-dependent, delays between impacts and QPE peak of emission. In the consistency check below, we then use as upper and lower  limits on $P_{\rm mod}$  and $A_{\rm mod}$ those obtained by assuming a $5$\% uncertainty on the best-fitting parameters whenever the statistical ones are smaller.

From the O-C analysis of GSN~069 (see Table~\ref{tab:OCgsnmix}), we derive two possible sets of $P_{\rm mod}$ and $A_{\rm mod}$. By inserting their upper and lower limits into Eq.~\ref{eq:Kep}, $M_{\rm tot}$ is consistent with the upper limit from the $M-\sigma$ relation ($\sim 3.2\times 10^6~M_\odot$) if $i_{\rm obs}\gtrsim 55^\circ$ and $q\lesssim 0.2$ for the $\sim 19$~d modulation, and $i_{\rm obs}\gtrsim 29^\circ$ for any $q\leq 1$ for the $\sim 43$-$44$~d one. Hence, there is significant room for the presence of a SMBH binary in GSN~069 in both cases.

We have modified the numerical code presented in \citet{2023A&A...675A.100F} introducing a second SMBH with mass $M_2$ that forms, with the inner EMRI, a SMBH outer binary  with orbital period $P_{\rm out}$, and computed the QPE times of arrival to assess whether a hierarchical triple system can account for the observed periodic modulation and correlation of the O-C diagrams in GSN~069 (see Appendix~\ref{sec:notesmodel} for a description of the numerical implementation).  We considered the cases of both a $P_{\rm out} = 44$~d and $19$~d assuming that the outer binary is responsible for the O-C modulation via light travel time effects. In both simulations, the outer SMBH binary orbit and the accretion disc around $M_1$ were assumed to lie in the $x$-$y$ plane, while $i_{\rm EMRI} = 10^\circ$ with respect to the $z$-axis. Our goal here was not to obtain an accurate fit to the data, but rather to investigate whether a SMBH outer binary could represent a viable qualitative solution for the observed QPE timing fulfilling the conditions outlined at the end of Section~\ref{sec:wayforward}. Therefore, we did not vary the system geometry to search for a better quantitative agreement.

The first simulated system is composed by an inner, QPE-emitting EMRI with $M_1 = 8\times 10^5~M_\odot$ (and $m\ll M_1$), eccentricity $e =0.05$, and orbital period $\sim 18$~hr (see Table~\ref{tab:OCgsnmix}) and a second SMBH with $M_2=2\times 10^6~M_\odot$. The total nuclear mass (ignoring the EMRI secondary) was then $M_{\rm tot} = 2.8\times 10^6~M_\odot$, consistent with the range inferred from the $M-\sigma$ relation in GSN~069 (see Section~\ref{sec:BHmasses}). The outer binary was set on a circular orbit with orbital period $P_{\rm out} =44$~d. The observer inclination was set to $i_{\rm obs} = 60^\circ$. The resulting outer SMBH binary semi-major axis is $\sim 1.67\times 10^{-4}$~pc, and the triplet is stable with a merger time of $\sim 0.1$~Myr for the outer SMBH binary. The second simulation was realised with $M_1 =2.2\times 10^5~M_\odot$, $M_2=2.8\times 10^6~M_\odot$, $P_{\rm out} = 19$~d, and $i_{\rm obs} = 75^\circ$. The resulting outer SMBH binary semi-major axis is $\sim 9.8\times 10^{-5}$~pc, and the triplet is stable with a relatively short merger time of $\sim 2.8\times 10^4$~yr for the outer SMBH binary. 

\begin{figure}[t] 
    \centering
        \includegraphics[width=0.9\columnwidth]{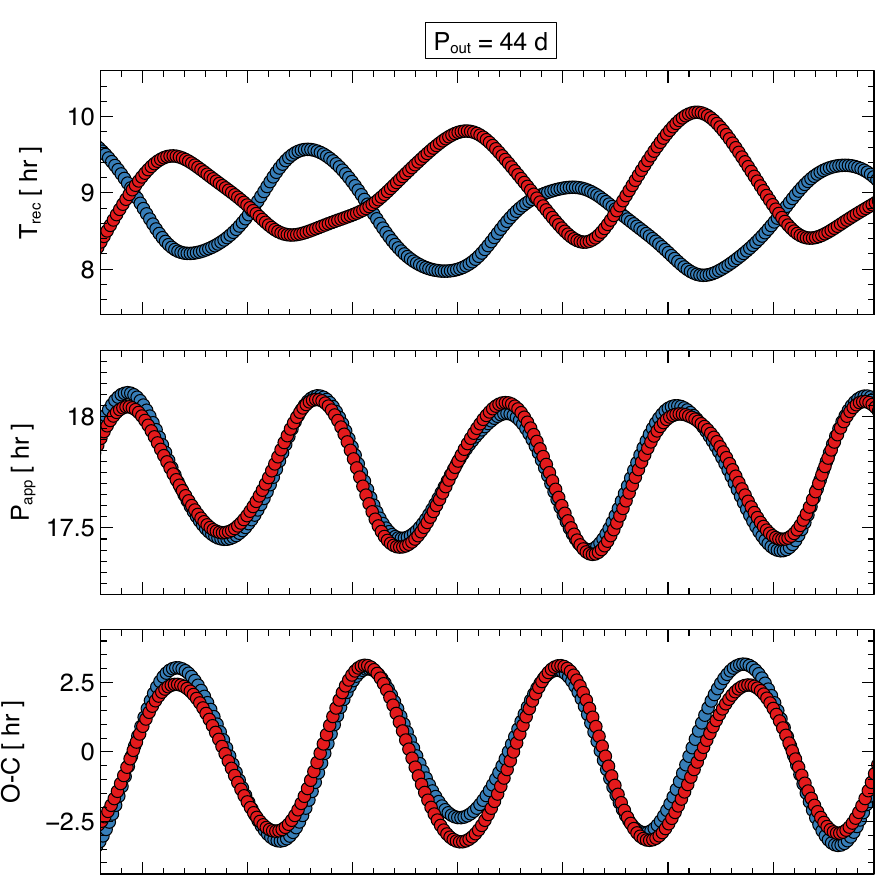}
\includegraphics[width=0.9\columnwidth]{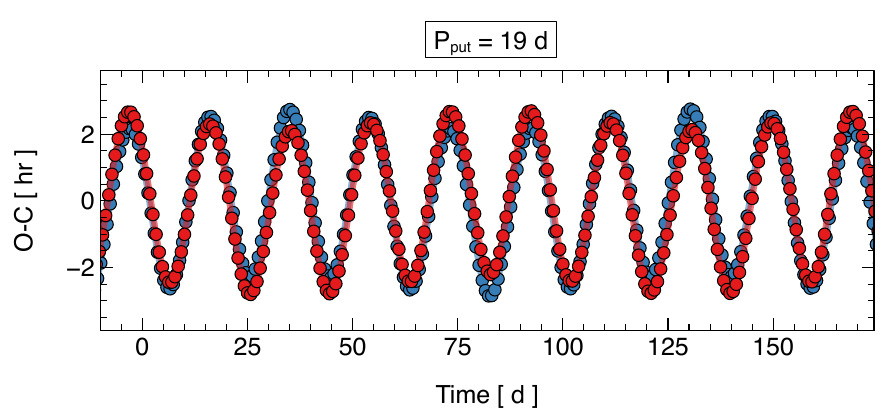}%        
    \caption{Hierarchical triple solution for GSN~069. In the upper three panels we show $T_{\rm rec}$, $P_{\rm app}$, and the O-C diagrams for a hierarchical triple solution for GSN~069 comprising the inner, QPE-emitting EMRI and an outer circular SMBH binary with orbital period $P_{\rm out} =44$~d. The O-C diagrams for the alternative solution with $P_{\rm out}=19$~d are shown in the lower panel.}
    \label{fig:GSNtriple}
\end{figure}

In the upper three panels of Fig.~\ref{fig:GSNtriple}, we show the resulting $T_{\rm rec}$, $P_{\rm app}$, and O-C diagrams for $P_{\rm out} =44$~d while, in the lower panel, we only show the O-C diagrams for $P_{\rm out} = 19$~d. The simulated quantities satisfy all properties outlined at the end of Section~\ref{sec:wayforward}, and thus qualify as a viable solution for the QPE timing behaviour of GSN~069: the alternating recurrence times are preserved, but the two branches are not exactly anti-correlated allowing, for instance, for time intervals in which the drop (rise) in one branch is more pronounced than the rise (drop) in the other, as observed in Fig.~\ref{fig:gsn}. The odd and even branches in both $P_{\rm app}$ and the O-C diagrams are well correlated, roughly in phase, and periodic at the outer binary orbital period $P_{\rm out}$. In particular, the simulated O-C diagrams (two lower panels in Fig.~\ref{fig:GSNtriple}) can be compared to the corresponding observed ones in  Fig.~\ref{fig:GSNOCmix} as well as with those associated with different QPE identifications in Fig.~\ref{fig:GSNOClin} and \ref{fig:GSNOCquad}. The agreement between the observed and simulated O-C diagrams in timescale, shape, and modulating amplitude is excellent.

We conclude that a hierarchical triple system composed by the inner, QPE-emitting EMRI and an outer SMBH binary with sub-milliparsec separation is a viable solution that can account for the observed periodicity and correlation of the O-C diagrams for odd and even QPEs while preserving the alternating recurrence times and aligning the orbital timescale $P_{\rm app}$ for the two branches on the same functional form. 

\section{Quiescent X-ray emission modulation}
\label{sec:modulation}

In principle, both the disc precession and hierarchical triple scenarios outlined above might induce a modulation of the quiescent (out-of-QPEs) X-ray disc emission in GSN~069 on the same timescale over which the O-C diagrams are modulated. The variability of the quiescent X-ray emission in GSN~069 can thus be used to confirm the external modulation scenario suggested by the O-C analysis, and perhaps even to constrain its origin. 

Disc precession generally modulates the disc emission on the precession timescale, but it is not necessarily associated with an X-ray modulation. This is because, the modulation probed via the O-C analysis is associated with delayed or anticipated QPEs that are produced by the impacts between the EMRI's secondary and the disc. Such impacts occur on a ring on the disc whose inner and outer boundaries are set by the pericentre and apocentre distances of the EMRI orbit, i.e. $a\,(1\pm e)$ where $a$ and $e$ are the orbit semi-major axis and eccentricity. Given the assumed parameters in GSN~069 ($P_{\rm orb} \simeq 18$~hr, $e_{\rm orb} \simeq 0.05$, and $M=8\times 10^5~M_\odot$), impacts occur on a ring at $180$-$200~R_g$ from the centre. This portion of the disc is significantly farther away than the X-ray emitting region (likely few tens of $R_g$ only), and it is instead associated with UV or optical disc emission, so that the disc precession model predicts periodic variability at these longer wavelengths. This is currently difficult to probe due to stellar contamination in the galactic nucleus and monitoring \hst\ observations over a tens of days baseline are needed to explore this possibility. 

\begin{figure*}[t] 
    \centering
        \includegraphics[width=1.8\columnwidth]{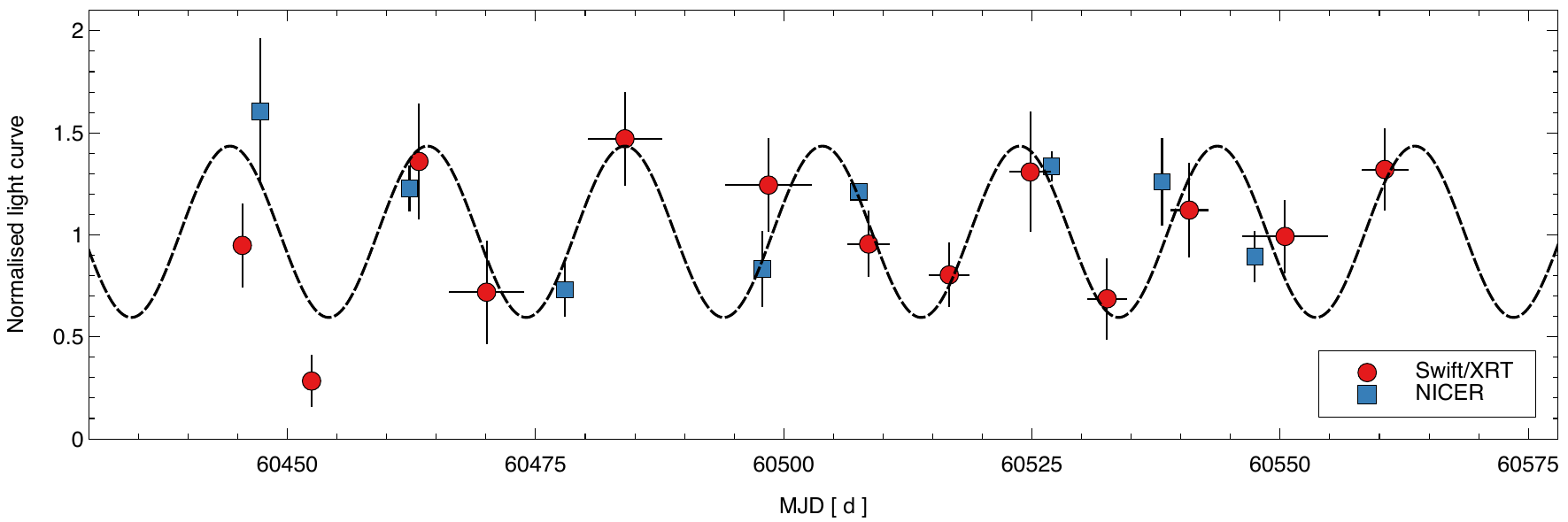}
    \caption{2024 \swift\ and \nicer\ monitoring campaign. We show the normalised 0.3-1~keV quiescent light curves of GSN~069 as obtained by the \swift~XRT and by NICER during the current campaign. Any two consecutive Swift observations with consistent count rates have been combined to reduce the statistical uncertainty. Individual NICER snapshots have been analysed separately, and the data points represent the mean 0.3-1~keV flux and standard deviation obtained within a few hours. The dashed line is a sine function modulation with period $\simeq 19.9$~d and semi-amplitude $\simeq 45$\% resulting in $\chi^2 =31$ for $17$ degrees of freedom.}
    \label{fig:monitoring}
\end{figure*}

X-ray periodic variability of the quiescent disc emission is only expected if the disc precesses rigidly so that the inner X-ray emitting disc also precesses on the same timescale. This is not granted, as the disc might not fulfil the conditions for rigid precession \citep{2016MNRAS.455.1946F}, or might break or tear into rings precessing on their own timescale. The precessing disc is bound to align with time on a  timescale that can be of the order of years for relatively low black hole mass and  effective viscosity values, so that observing a precessing disc in GSN~069 years after the TDE-like outburst is possible \citep{2016MNRAS.455.1946F}. Moreover, if rigid disc precession holds, viscous spreading of the disc with time leads to an increase of the disc precession timescale during alignment, a prediction that might be tested if the quiescent X-ray emission is indeed modulated. As mentioned in Section~\ref{sec:discprecession}, a good match with the O-C modulation amplitude is obtained for a disc misalignment of the order of $20^\circ$. The predicted X-ray flux modulation is however also a function of the observer inclination $i_{\rm obs}$. Since $i_{\rm obs}$ has a minor effect on O-C data, we could not constrain it, and the expected X-ray flux modulation amplitude  in the case of (rigid) disc precession spans a wide range that corresponds to the different possible lines-of-sight.

In the case of a triple system (the inner binary containing the EMRI and an outer SMBH), orbital motion of the EMRI system about the centre of mass does not only modulate QPEs times of arrival, but also the X-ray emission from the inner regions of the accretion disc via Doppler boosting. Assuming that the emitted radiation has spectrum $F\propto \nu~^\alpha$ in frequency space, and for orbital velocities $\beta_{\rm orb} = v_{\rm orb}/ c \ll 1$, the fractional variability due to Doppler boosting from the motion within a circular binary can be written as
\begin{align} 
\frac{\Delta F}{F} = \left( 3-\alpha \right)~\beta_{\rm orb}~\sin i_{\rm obs}~\cos \phi_{\rm orb}\,, \label{eq:doppler}
\end{align}
where $\phi_{\rm orb}$ is the orbital phase, and $i_{\rm obs}$ is the observer inclination  \citep{2003ApJ...588L.117L,2015Natur.525..351D,2018MNRAS.476.4617C}. 

However, as discussed in Section~\ref{sec:triples}, when interpreting the O-C modulation in terms of time delays due to orbital motion in an SMBH binary, the O-C modulation amplitude $A_{\rm mod}$ sets the radius of the orbit of the EMRI system around the centre of mass with the outer SMBH as $R = A_{\rm mod}~c/ \sin i_{\rm obs}$, while the period of the O-C modulation ($P_{\rm mod}$) is assumed to match the SMBH binary orbital period, i.e.  $P_{\rm mod} = P_{\rm out} $.  Therefore, the orbital velocity is simply 
\begin{align} 
v_{\rm orb} & = 2\pi~\frac{R}{P_{\rm out}} = 2\pi~\frac{A_{\rm mod}~c}{\sin i_{\rm obs}~P_{\rm mod}}\,.
\end{align}
By inserting the orbital velocity into Eq.~\ref{eq:doppler}, we obtain 
\begin{align} 
\frac{\Delta F}{F} = 2\pi~\left( 3-\alpha \right)~\frac{A_{\rm mod}}{P_{\rm mod}}~\cos \phi_{\rm orb}\,, \label{eq:dopplermax}
\end{align}
that can be used to estimate the predicted modulation semi-amplitude, reached for $\cos \phi_{\rm orb} = 1$ corresponding to the orbital phase when the X-ray emitting disc (and the EMRI binary) are on the approaching side of the outer binary orbit. By considering that the thermal SED of GSN~069 has typical spectral index $\alpha \simeq -9$ in the 0.3-1~keV band, and inserting the numbers for $A_{\rm mod}$ and $P_{\rm mod}$ obtained from the O-C analysis, the fractional variability semi-amplitudes in the 0.3-1~keV X-ray band  is then 
\begin{align} 
& \left(\frac{\Delta F}{F}\right)_{\rm 19~d}   \simeq 42\% & \mathrm{or} &
& \left(\frac{\Delta F}{F}\right)_{\rm 43-44~d}   \simeq 20\% \label{eq:doppler19}\,,
\end{align}
where we have assumed the mean $A_{\rm mod}$ and $P_{\rm mod}$ from Table~\ref{tab:OCgsnmix} for the two possible periods. Naturally, the shortest period corresponds to the highest-amplitude modulation since the orbital velocity is the highest, thus maximising the effect of the Doppler boost. Hence, if the O-C modulation is due to light-travel-time effects in an outer SMBH binary, the Doppler boosting model predicts the X-ray flux variability amplitude with no free parameters, as the amplitude only depends on measured quantities (spectral index $\alpha$, and the amplitude $A_{\rm mod}$ and period $P_{\rm mod}$ of the O-C modulation). This clear prediction can then be tested against monitoring X-ray data.

GSN~069 was monitored on several occasions in the X-rays precisely in the attempt to search for a periodic modulation of the quiescent disc emission. However, the quiescent X-ray variability on both short and long timescales had typically an amplitude well above the $50$\% level, which prevented us from looking for relatively low-amplitude modulations (results will be presented elsewhere). \swift\ and \nicer\ are currently monitoring GSN~069 over an extended baseline, and the source has apparently entered a period of high average flux and relatively low-amplitude X-ray variability since May 2024. We note that, consistent with the QPE disappearance at high fluxes reported by \citet{2023A&A...674L...1M}, no QPEs have been detected so far in the on-going campaign since May 2024 which also comprises a long-enough $\sim 120$~ks \xmm\ observation that failed to detect any clear QPE. 

The current X-ray light curve from \swift\ and \nicer\  is shown in Fig.~\ref{fig:monitoring}. In both cases, the light curves have been normalised to the respective best-fitting constant model during the campaign to remove calibration uncertainties between detectors, and to ease comparison as we are here interested in fractional variability amplitudes. The \swift\ XRT typically collects few tens of counts per observation so that no spectral information is available. Hence, the data points in Fig~\ref{fig:monitoring} represent the normalised 0.3-1~keV count rate. Any two consecutive \swift\ observations delivering consistent count rates have been combined to improve the signal-to-noise. On the other hand, each \nicer\ data point comprises a series of snapshot exposures with typical duration of few hundred seconds. We have analysed the X-ray spectra of each individual snapshot exposure\footnote{This is necessary to account for the variable \nicer\ background, see Appendix~\ref{sec:data} for details.}, and the resulting X-ray flux from exposures within a few hours has then been combined. Each \nicer\ data point in Fig.~\ref{fig:monitoring} represents the average 0.3-1~keV X-ray flux (and standard deviation) from exposures within a few hours, normalised to the best-fitting constant model throughout the \nicer\ campaign. 

A constant model fit to the combined \swift\ and \nicer\ data sets shown in Fig.~\ref{fig:monitoring} results in $\chi^2 = 95.0$ for $20$ degrees of freedom. In order to search for possible modulations, we added a sinusoidal function (three free parameters) to the model which resulted  $\chi^2 =31$ for $17$ degrees of freedom, i.e. an improvement that is statistically significant at the $\sim 99.98$\% level from a simple F-test.  The corresponding, best-fitting sinusoidal modulation is shown as dashed line in Fig.~\ref{fig:monitoring}. Although the statistical improvement obtained by adding the sinusoidal modulation to a constant model is formally high, the number of cycles is still too small to draw firm conclusions about the presence of the X-ray modulation, considering also the sparse sampling in Fig.~\ref{fig:monitoring}.

The best fitting period and semi-amplitude are $P = 19.9\pm 0.3$~d and $A = 0.45\pm 0.06$.   The modulating period is therefore consistent, within $3$-$4$\%, with the $\sim 19$~d modulation inferred from the completely independent technique of O-C diagrams, which is based exclusively on QPEs times of arrival, while the (tentative) flux modulation from the \swift\ and \nicer\ data is obtained from out-of-QPEs quiescent flux variability. Remarkably, the variability amplitude is consistent with that predicted from Doppler boosting for the $\sim 19$~d period ($\simeq 42$\%, see Eq.\ref{eq:doppler19}). In the case of a rigidly precessing disc, the $40$-$50$\% modulation suggested by the \swift\ and \nicer\ data can be reproduced by various combinations of disc misalignment and observer inclination although, for consistency with the O-C modulating amplitude, a disc misalignment of the order of $20^\circ$ appears favoured (see Section~\ref{sec:discprecession}).

\begin{figure}[t] 
    \centering
        \includegraphics[width=0.9\columnwidth]{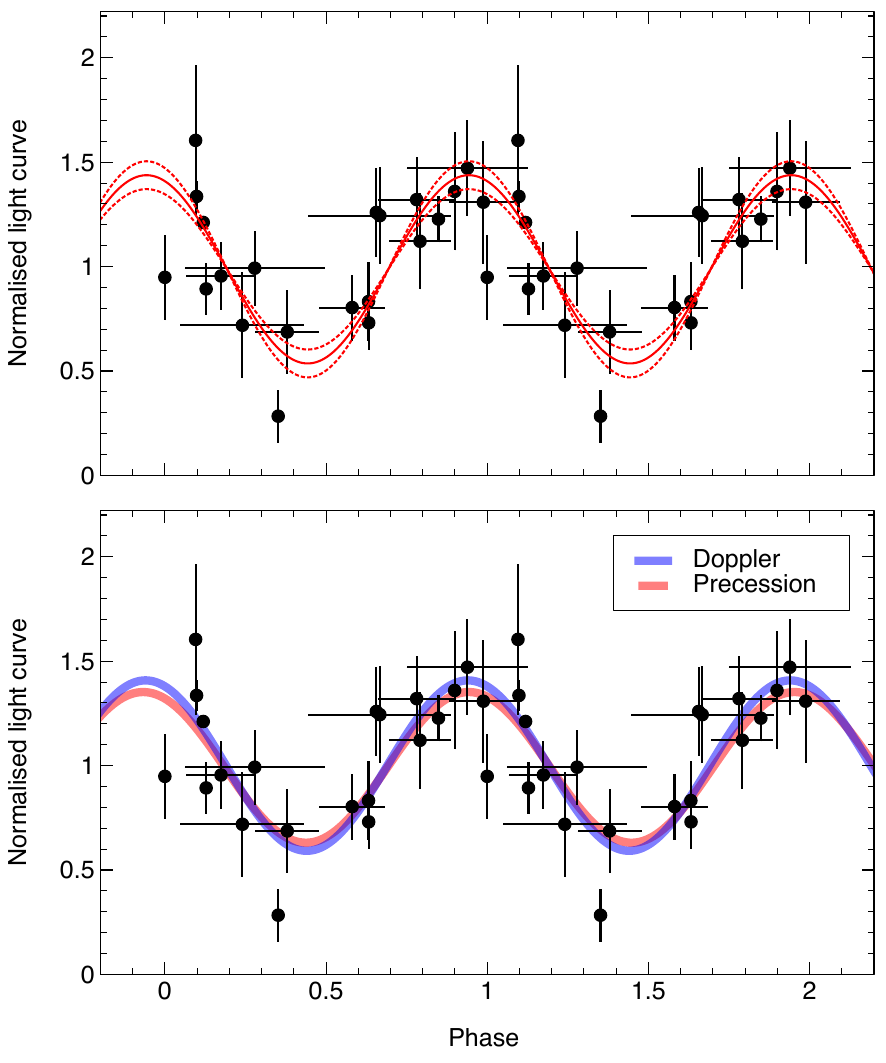}
    \caption{Folded \swift\ and \nicer\ light curve. The upper panel shows the light curve of Fig.~\ref{fig:monitoring} folded at the best-fitting period of $19.9$~d. Two cycles are shown for visual clarity, and the distinction between \swift\ and \nicer\ data points has been removed. The solid red line represents the sinusoidal modulation with $1\sigma$ uncertainties in its amplitude shown as dotted lines. In the lower panel, we show the predicted modulation from Doppler boosting (blue) as well as one possible realisation of the disc precession model (red) corresponding to a disc misalignment of $i_{\rm disc} = 20^\circ$ and observer inclination $i_{\rm obs} = 45^\circ$.}
    \label{fig:folded}
\end{figure}

In the upper panel Fig.~\ref{fig:folded}, we show the normalised light curve folded at the best-fitting period of $19.9$~d with the sinusoidal best-fitting modulation over-plotted to guide the eye as a solid line (dotted lines represent the $1\sigma$ range of allowed amplitude). In the lower panel, we show the same data together with the expected Doppler boosting modulation (blue), and one realisation of the disc precession induced modulation in which we consider a disc misaligned by $20^\circ$ with respect to the SMBH's spin (consistent with the O-C modulation), and an intermediate observer inclination of $45^\circ$. 

\begin{figure}[t] 
    \centering
        \includegraphics[width=0.9\columnwidth]{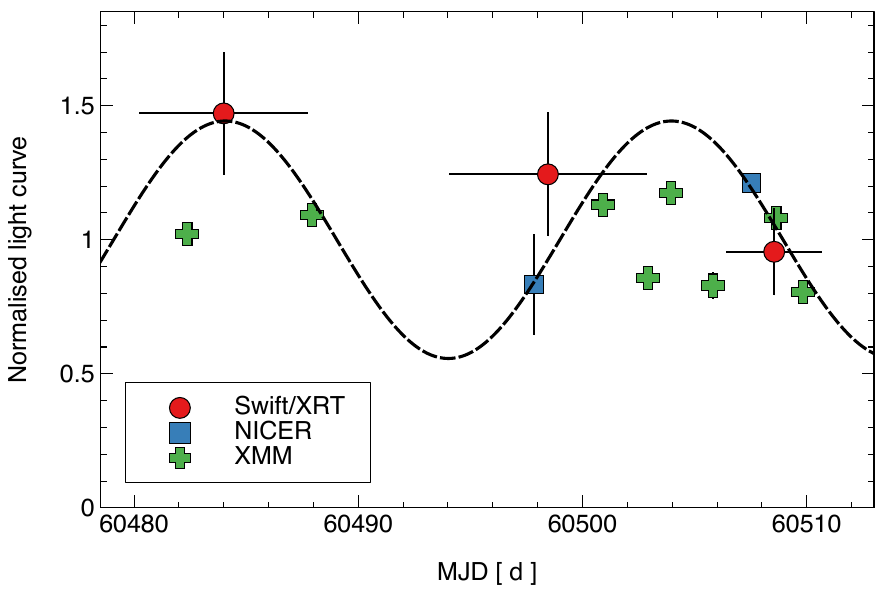}
    \caption{Zoom of Fig.~\ref{fig:monitoring} over a restricted MJD range in which eight \xmm\ observations are also available (green crosses). The \xmm\ data points show the 0.3-1~keV average X-ray flux during each \xmm\ observation, normalised to the best-fitting constant model over the spanned baseline. Uncertainties on the \xmm\ data are included, but smaller than symbol size.}
    \label{fig:xmm24}
\end{figure}

However, a series of eight \xmm\ observations obtained between MJD $60\,482$ and $60\,510$ do not fully confirm the modulation tentatively seen with \swift\ and \nicer. Only a few (three or four) \xmm\ observation show quiescent 0.3-1~keV fluxes roughly consistent with the modulation shown in Fig.~\ref{fig:monitoring}, while others have lower-than-predicted flux level by a significant amount. This is shown in Fig.~\ref{fig:xmm24} where we reproduce the light curve of Fig.~\ref{fig:monitoring} in a restricted range and included the normalised 0.3-1~keV flux resulting from spectral analysis of the \xmm\ EPIC-pn camera's data (see Appendix~\ref{sec:data} for details). The discrepancies between the \xmm\ data and the sinusoidal modulation, suggested by the \swift\ and \nicer\ monitoring, casts some doubts on the reality of the quiescent flux periodicity, and only further higher cadence, monitoring observations on a longer baseline can provide a firm result, clarifying whether the discrepant \xmm\ data points can be attributed to occasional intrinsic variability. 

We are continuing to monitor GSN~069 with \swift\  every few days, and we have significantly increased the cadence of the \nicer\ observations. The combined monitoring will enable us to assess the statistical significance of any modulation on firmer statistical grounds than possible at present. Having monitored the source also in the past, we must stress that there is no guarantee that the current relatively low-amplitude variability continues in the future. Past data show that the X-ray flux enters phases or relatively low intrinsic variability that are reminiscent of sinusoidal behaviour similar to that shown in Fig.~\ref{fig:monitoring}, but these phases are often interrupted by periods of enhanced variability during which it is difficult, if not impossible, to search for lower-amplitude modulations. At this stage, we refrain from providing possible explanations for the observed discrepancy between the \xmm\ and the \swift\ and \nicer\ measurements as they would represent mere speculations. Future monitoring data are the only reliable way of assessing whether the quiescent disc emission in GSN~069 is indeed modulated periodically.

\section{Discussion}
\label{sec:discussion}

Within the context of the impacts model for QPEs, a consistent picture is starting to emerge from observations. The data are consistent so far with pre-existing EMRI systems that are revealed electromagnetically through QPEs whenever an accretion disc extending further than the secondary orbit is present. Interactions between the disc and the orbiting secondary produce the observed high-amplitude quasi-periodic flares. The association between QPEs and TDEs, first suggested in the case of GSN~069 \citep{2018ApJ...857L..16S,2019Natur.573..381M} and then strengthened by the detection of QPEs in an optically-selected TDEs \citep{2023A&A...675A.152Q,2024Natur.634..804N} among other evidences, indicates that the accretion disc is likely the result of the full or partial tidal disruption of a star by an otherwise inactive SMBH. Such a disc is initially compact, with outer radius of the order of few times the tidal radius (typically tens of $R_g$), and could be smaller than the typical EMRI orbit. As the accretion disc spreads out viscously, it intercepts the EMRI orbit and gives rise to the start of the QPE phenomenon. QPEs are therefore generally expected to only appear at late times, with a delay with respect to the TDE peak. On the other hand, if the tidally disrupted star is evolved, as might be the case in GSN~069 \citep{2021ApJ...920L..25S}, its tidal radius can be of the order of hundreds of $R_g$, so that QPEs due to EMRI orbits with semi-major axis of $100$-$200~R_g$ may have started immediately. In GSN~069, a sufficiently long X-ray observation $\sim 4.5$~yr after the initial X-ray outburst failed to detect any QPEs. These were first detected in December 2018, about $8.5$~yr after the X-ray peak. Hence, QPEs in GSN~069 appeared with a delay of $\sim 4.5$-$8.5$~yr with respect to the initial TDE-like outburst, which may be taken as an indication of a delayed QPE appearance. However, later observations have shown that QPEs disappear above a given X-ray luminosity (or mass accretion rate) threshold which may be the reason why no QPEs were detected during the sufficiently long but high flux observation $\sim 4.5$~yr after TDE peak  \citep{2023A&A...674L...1M}. 

\subsection{QPE timing properties: disc precession or a SMBH binary}
\label{sec:discussiontiming}

The impacts model for QPEs, in which the secondary EMRI component pierces through a geometrically thin, non-precessing accretion disc around the primary, produces alternating recurrence times for any non-zero orbital eccentricity, in good agreement with the observed QPE timing properties in GSN~069 and other QPE sources. Such a model also naturally implies that odd and even QPEs share a common baseline period (and period derivative, if present), as this is set by the EMRI orbital properties and evolution. However, when no external modulation is included, the model predicts that recurrence times $T_{\rm rec}$ (separation between consecutive QPEs), apparent orbital period $P_{\rm app}$ (separation between consecutive QPEs of the same parity), and O-C diagrams for odd and even QPEs are periodic on the (super-orbital) apsidal precession timescale and, most importantly, are all in phase opposition. In absence of an external modulation that is faster than apsidal precession and of sufficiently high amplitude to dominate the timing, these are unavoidable consequences of relativistic orbits dynamics. 

None of the three observed quantities in GSN~069 appear to comply with this very clear expectation. In particular, the O-C diagrams are consistent with being periodic on a super-orbital timescale of the order of tens of days, but odd and even QPEs data are well correlated with minimal phase difference (actually simply due to the fact that odd and even QPEs are not simultaneous but separated by $T_{\rm rec}$), contrary to model's predictions.  Also, the amplitude of the O-C modulation is about one order of magnitude higher than that expected from apsidal precession alone, given the modest EMRI's eccentricity inferred in GSN~069 ($\lesssim 0.15$). On the other hand, our O-C analysis shows that solutions in which odd and even QPEs  share a common period (and, if present, period derivative) do exist, which is fully consistent with the impacts scenario, and suggests that the model might need to be modified rather than rejected. 

The observed modulation of the O-C diagrams in GSN~069 is likely associated with an external physical mechanism unrelated to the EMRI orbital properties. We have explored two possible sources of modulation: disc precession and light-travel-time effects due to the motion of the QPE-emitting system (EMRI plus disc) around the centre of mass of an outer SMBH binary system. Both mechanisms qualitatively reproduce the correlated modulation (with period $P_{\rm mod}$) of the O-C diagrams for odd and even QPEs, where $P_{\rm mod}$ is set by the disc precession timescale or the outer SMBH binary orbital period. Fig.~\ref{fig:r2} illustrates the effects on O-C diagrams (or QPEs time delays with respect to a given reference) of the various processes considered in this work. 

As shown in Fig.~\ref{fig:GSNprecession} and \ref{fig:GSNtriple}, the two driving mechanisms produce almost identical O-C diagrams, so that this technique is not very useful to pinpoint the modulating physical process at play. However, the two scenarios can be easily distinguished from the shape of $P_{\rm app}$ (that is the time intervals between consecutive QPEs of the same parity) over at least one cycle. In the case of disc precession, $P_{\rm app}$ exhibits distinctive peaks\footnote{Note than if the disc and EMRI's orbit angular momenta do not have the same sign (with respect to the SMBH spin), the peaks in $P_{\rm app}$ revert into valleys.} repeating every disc precession period and separated by intervals of nearly constant behaviour. On the other hand, in the case of a hierarchical triple system, the $P_{\rm app}$ evolution is sinusoidal-like with period equal to that of the outer binary (see Fig.~\ref{fig:GSNprecession} and \ref{fig:GSNtriple}). A sufficiently dense X-ray monitoring campaign over at least one cycle in which the individual continuous exposure is long enough to detect at least three consecutive QPEs will enable us to accurately measure the shape of $P_{\rm app}$, and therefore to distinguish between the two proposed scenarios.

Another possible way of distinguishing between the two physical processes is through a new O-C campaign on GSN~069. In the case of a SMBH binary, the modulating period decay must be consistent with that from GW emission. Assuming fiducial parameters for a circular SMBH binary with total mass of $3\times 10^6$, mass ratio $q=0.1$, and orbital period of $\simeq 19$~d (similar to that providing a good match with the data), the GW period derivative is $\dot P_{\rm GW} \simeq 8.3\times 10^{-7}$ or $\sim 3\times 10^{-4}$~d per year, which is undetectable in the X-rays even on a 10 or 20 years baseline. The same is true for the modulating amplitude because the outer SMBH binary orbit does not harden sufficiently fast to lead to detectable amplitude decay. In the case of disc precession, the disc precession timescale is expected to be longer in the future as the disc extends farther out due to viscous spreading. Moreover, as the disc alignment process continues, the modulating amplitude is also expected to decay on years timescales. A new O-C campaign detecting a longer modulating period with lower amplitude would therefore favour the disc precession model and rule out the SMBH binary one.    

\subsection{Hints for period decay in GSN~069 and possible interpretations}

The ambiguity on QPE's identification during the May 2019 observation of GSN~069 implies that the EMRI orbital period $P_{\rm orb}$ might be decaying. The O-C data are consistent with three possible solutions for the period derivative, namely $\dot P_{\rm orb} = 0$, $-3$-$4\times 10^{-5}$, or $-6$-$7\times 10^{-5}$. An hint for a period decay of $\dot P_{\rm orb} \simeq -3$-$4\times 10^{-5}$, consistent with the O-C solutions discussed in Section~\ref{sec:gsnOC}, is obtained by comparing $P_{\rm orb}$, as derived from the O-C data between December 2018 and May 2019, with the average $P_{\rm app}$ (a proxy for the actual $P_{\rm orb}$) in a June-July 2023 campaign. This estimate cannot be considered as a secure detection because it is only based on the average $P_{\rm app}$ in 2023, which may significantly underestimate the actual EMRI orbital period due to the restricted number of independent measurements in 2023.  However, with the caveat that the detection of such a $\dot P$ is tentative only, we discuss its possible implications in some more detail. 

One obvious source of period decay in an EMRI system is gravitational wave (GW) emission. However, if the EMRI secondary is a star rather than a black hole, hydrodynamical drag from repeated star-disc collisions naturally induces a (negative) period derivative as well \citep[for relevant formulae, see e.g.][]{2023ApJ...957...34L,2024MNRAS.527.4317L,2024A&A...690A..80A}. Assuming that the tentative $\dot P_{\rm orb}$ is only due to GW emission, its amplitude together with the derived EMRI orbital period and eccentricity implies that, even for a primary black hole mass at the high end of the range allowed from the $M-\sigma$ relation ($\simeq 3.2\times 10^6~M_\odot$), the secondary mass needs to be $m\gtrsim 10^5~M_\odot$ effectively defining a SMBH binary rather than an EMRI (semantically defined as a binary with mass ratio $q\leq 10^{-4}$). 

If the period decay is instead due to hydrodynamical gas drag, $\dot P_{\rm drag} \approx - 3\pi~R^2_\star~\Sigma~m^{-1}_\star$ where $R_\star$ and $m_\star$ are the radius and mass of the orbiting star and $\Sigma$ is the disc surface density \citep{2024MNRAS.527.4317L,2024A&A...690A..80A}. For a Sun-like star that does not change structure due to collisions, a relatively high $\Sigma \gtrsim 10^6$~g~cm$^{-2}$ at the impacts site would be required to match the observed period decay in GSN~069. Considering fiducial parameters for GSN~069 (a non-spinning SMBH with mass of  $8\times 10^5~M_\odot$, secondary orbital period of $18~hr$, and an average impacts radius of $\simeq 190~R_g$) and assuming a surface density radial profile $\Sigma=\Sigma_0(R/R_{\rm g})^{-p}$ with $p=3/5$, the condition $\Sigma \gtrsim 10^6$~g~cm$^{-2}$ at impact implies an overall accretion disc mass $M_{\rm disc} \gtrsim 1.2~M_\odot$ for a disc extending out to $\approx 200R_{\rm g}$, the minimum outer radius that allows for two impacts per orbit\footnote{Note that \cite{2023A&A...675A.100F} assumed a $4M_{\odot}$ disc for GSN~069 when reproducing single-epoch QPE light curves with their impacts plus disc precession model.}. Since about $50$\% of the tidally disrupted debris end up forming the accretion disc, such a disc mass implies the \citep[likely partial, see][]{2023A&A...670A..93M,2024ApJ...961L...2B} disruption of a star, or envelope, with mass $\gtrsim 2.4~M_\odot$. 

We point out, however, that the repeated collisions between the star and the disc would generally result in the inflation of the star's outer layers \citep{2024arXiv240714578Y}. Since, in the hydrodynamical gas drag scenario,  the disc surface density scales as $\Sigma \propto \dot{P}_{\rm orb}~m_\star~R_\star^{-2}$, a lighter star whose radius was inflated above the equivalent main-sequence value would result in a lower inferred $\Sigma$, possibly down to $\sim 10^4$~g~cm$^{-2}$ at the collisions site, thus leading to a significantly lower minimum disc mass.  As mentioned in Section~\ref{sec:gsnOC}, other mechanisms related to longer-term modulations (e.g. nodal precession of the EMRI orbital plane in case of a Kerr primary SMBH) might give rise to apparent $\dot P_{\rm orb} \neq 0$ or, at least, to more complex behaviour as discussed, for instance, by \citet{2024A&A...690A..80A}. This suggests that the tentative $\dot P_{\rm orb}$ we report should not be over-interpreted.

\subsection{Modulation of the quiescent disc emission}

As discussed in Section~\ref{sec:modulation}, \swift\ and \nicer\ monitoring observations from May to September 2024 indicate a possible modulation of the X-ray disc emission in GSN~069 with a period of $\sim 19.9$~d and semi-amplitude of $40$-$50$\%. The period is formally slightly longer than the $\sim 19$~d solution of the O-C analysis from QPEs times of arrival but consistent with it within a few per cent which, considering the likely systematic uncertainties affecting the O-C results, suggests not to claim a statistically significant difference at this stage. The relatively small number of observed cycles, sparse sampling, and some discrepant \xmm\ data points, suggest to consider the X-ray flux modulation as tentative only. Nevertheless, it is encouraging that the tentative period is broadly consistent with the O-C solution associated with the $\sim 19$~d modulation, as the two analysis are completely independent from one another, which provides some support to the robustness of both.    

 Disc precession can be associated with a wide range of possible modulating amplitudes, depending on the actual disc misalignment with respect to the SMBH spin axis (here simply the $z$-axis) and on observer inclination. The X-ray flux modulation amplitude is consistent with that expected from a rigidly precessing accretion disc for high observer inclination (close to edge-on) if the disc is only slightly misaligned or for low observer inclination and large disc misalignment. A good match with the O-C modulation amplitude of $\sim 2.5$~hr is reached for a disc misalignment of $\simeq 20^\circ$ with repect to the $z$-zxis. Assuming this misalignment, the quiescent X-ray flux modulation is consistent with intermediate observer inclinations of the order of $45^\circ$. 
 
In the case of Doppler boosting of the inner disc emission in an outer SMBH binary, the predicted variability amplitude has no free parameters once a solution for the O-C analysis is selected (with either $P_{\rm mod}\simeq 19$~d or $\simeq 43$-$44$~d). This is because, if the O-C diagrams modulation is interpreted within this context, the orbital velocity is set by the orbit's size and period, that are measured through $A_{\rm mod}$ and $P_{\rm mod}$ modulo the sine of observer inclination, which however cancels out in Eq.~\ref{eq:doppler}. Considering the observed steep X-ray spectrum, Doppler boosting predicts a relatively precise $\sim 42$\% variability amplitude for the $\sim 19$~d O-C modulation, which is indeed consistent with the observed quiescent X-ray flux modulation of $40$-$50$\%. 

As discussed in Section~\ref{sec:discussiontiming}, the two scenarios make different predictions on the modulating period and amplitude time evolution, with only the disc precession model giving rise to detectable changes. Future monitoring observations able to follow the quiescent flux evolution in terms of both period and amplitude might then unveil the actual modulating process. As a further note, we point out that, in the SMBH binary scenario, QPEs time delays reach their maximum and minimum value at orbital phases corresponding to the far and near sides of the outer SMBH binary orbit with respect to the observer, while the extremes of the disc X-ray emission by Doppler boosting are expected at phases corresponding to the approaching and receding sides. A phase shift of $1/4~P_{\rm out}$ between the O-C delays and the X-ray flux variability is therefore expected. Detecting the Doppler-induced modulation of the X-ray flux at the correct phase with respect to O-C time delays would represent a smoking gun for the presence of a tight SMBH binary in the nucleus of GSN~069. This is not presently possible as the 2018-2019 observations (from which we derive QPEs delays) and the 2024 monitoring campaign are too far apart to be aligned considering the current period uncertainty of few per cent. However, future campaigns in which a sufficiently large number of QPEs are obtained quasi-simultaneously with quiescent X-ray fluxes on a few days cadence might prove key in this respect.

\section{Summary and conclusions}
\label{sec:conclusions}

We have studied the QPEs timing properties in GSN~069 primarily using X-ray data from three \xmm\ and one \cha\ observations between December 2018 and May 2019. These data were complemented by three \xmm\ observations in 2023, and by a series of \xmm, \swift, and \nicer\ monitoring observations between May and September 2024. Our primary goal was to compare the observed QPEs timing properties with predictions from one of the most popular models for QPEs, that of a secondary orbiting body piercing through the accretion disc around a primary SMBH in an EMRI system, with each impact producing one QPE. The main conclusions of our study are summarised below.

In 2018-2019, and with the caveats discussed in Section~\ref{sec:gsnOC}, O-C diagrams for odd and even QPEs are consistent with a periodic modulation on a timescale of either $\simeq 19$~d or $\simeq 43$-$44$~d with semi-amplitude of $\simeq 2.5$-$2.8$~hr. The sparse nature of the data prevents to distinguish between the two possible periods. The O-C data provide a measurement of the EMRI orbital period ($P_{\rm orb}\simeq 18$~hr) and are consistent with either no orbital evolution ($\dot P_{\rm orb} = 0$), or with period derivatives  $\dot P_{\rm orb} \simeq -3$-$4\times 10^{-5}$ or $\dot P_{\rm orb} \simeq -6$-$7\times 10^{-5}$, the intermediate value being supported by the estimated orbital period during a 2023 campaign, although we point out that an apparent period derivative might be induced by longer-timescale modulations that are currently out of observational reach ( see e.g. Fig.~\ref{fig:spin} for the case of orbital nodal precession induced by SMBH spin).

The O-C diagrams for odd and even QPEs have minimal phase difference (given by the average recurrence time between consecutive QPEs, $\sim 9$~hr or $\sim 0.38$~d). This is contrary to the expectations from the impacts model with no external modulation as O-C data for odd and even QPEs are predicted to be periodic on the apsidal precession timescale and in phase opposition, an unavoidable consequence of relativistic orbits dynamics. Moreover, the observed O-C modulating amplitude is about one order of magnitude higher than that expected from apsidal precession alone.

In order for the impacts model to apply, an external modulation, unrelated to the EMRI's orbit, needs to be considered. Through numerical simulations of impacts times between the secondary and the disc, we show that a rigidly precessing accretion disc or an outer SMBH - forming with the inner QPE-emitting EMRI a sub-milliparsec SMBH binary - qualitatively describe the QPEs timing data, reconciling the observed properties with the impacts model. Both models qualify as a qualitative solution for the QPE timing behaviour: they both preserve the alternating recurrence times while producing an evolution in which odd and even QPEs $T_{\rm rec}$ are not in phase opposition; they align the apparent orbital period $P_{\rm app}$ for the two branches on the same (or similar) function; they produce  O-C diagrams that are periodic on a super-orbital timescale (the disc precession or the outer SMBH orbital one) with minimal phase difference, rather than being in phase opposition.

We report evidence for a  periodic modulation of the quiescent X-ray (disc) flux in GSN~069 during a 2024 monitoring campaign with period $\simeq 19.9$~d and semi-amplitude of $\simeq 40$-$50$\%, although the relatively small number of cycles so far, as well as some discrepant data points suggest to consider the periodicity as tentative until (and if) further cycles are accumulated. The inferred period is consistent, within a few per cent, with the O-C analysis $\sim 5$~yr earlier for the case of the $\simeq 19$~d modulation. We point out that the two modulations are obtained from completely independent data and techniques supporting each other's robustness. Both scenarios proposed to explain the O-C modulation (disc precession and a SMBH binary) are consistent with the flux modulation amplitude. In fact, the SMBH binary model, in combination with results from the O-C analysis, predicts a flux variability of $42$\%, of the order of that tentatively observed, with no free parameters. Disc precession is consistent with the variability amplitude for a broad range of the two driving parameters, the disc's angular momentum misalignment and the observer inclination. A disc misalignment of $\simeq 20^\circ$, resulting in an O-C modulation amplitude consistent with the observed one, produces the observed X-ray flux modulation amplitude for intermediate observer inclination of $\sim 45^\circ$. The evolution of the flux modulation in both  period and amplitude might constrain the origin of the external modulation in the future, as the period is expected to increase (and the amplitude to decrease) only in the case of disc precession.

Both scenarios we propose, disc precession or the presence of a SMBH binary,  are plausible in TDEs and, given the mounting evidence that QPEs systems reside in TDEs, in  QPE systems. As the tidally disrupted star approaches the SMBH from a random direction, the bound debris will generally form a misaligned accretion disc with respect to the SMBH spin axis. As a result, disc precession is likely to take place initially, and could imprint multi-wavelength variability signatures \citep{2012PhRvL.108f1302S,2024Natur.630..325P}. If the disc is precessing rigidly, the innermost accretion disc emission is also modulated at the global disc precession timescale which depends on SMBH's mass and spin, but also on disc extent and structure. The alignment timescale depends on SMBH's and disc's properties, and it might last long enough to still be observable years after the TDE  as would be the case in GSN~069 \citep{2016MNRAS.455.1946F}.  On the other hand, the presence of a nuclear SMBH binary significantly enhances TDE rates in galactic nuclei due to the combined effect of the Kozai-Lidov secular mechanism \citep{1962AJ.....67..591K,1962P&SS....9..719L} and to resonant interactions of stars with the secondary black hole \citep{2005MNRAS.358.1361I,2011ApJ...729...13C}, so that finding SMBH binary candidates in TDEs is somewhat expected. In GSN~069, for stability reasons in a triple system, the EMRI plus disc QPE-emitting system needs to be associated with the lighter SMBH whose mass ratio with the more massive member of the SMBH binary is of the order of $q \lesssim 0.2$. Assuming, for example, the (non unique) parameters that provide a good match to the O-C data for the $\sim 19$~d solution, that is $M_{\rm tot} \simeq 3.02\times 10^6~M_\odot$, $q\simeq 0.079$, and $P_{\rm out} = 19$~d, the outer binary has a separation of $\simeq 9.8\times 10^{-5}$~pc or $\simeq 730~R_g$ (where the gravitational radius is that of the more massive SMBH), and it will merge in a relatively short time of the order of $2.8\times 10^4$~Myr.

Future X-ray monitoring observations, if appropriately designed, can be used to confirm (or reject) the overall impacts plus external modulation model as well as the reality of the quiescent (disc) X-ray flux periodicity. For the latter case, the on-going \swift\ and \nicer\ campaign has been extended, which should be sufficient to confirm the quiescent X-ray flux modulation, if present, within a few months. Moreover, future observations will enable us to distinguish between the two proposed scenarios by comparing their (different) predictions on QPEs timing and disc variability. Complementing such X-ray observations with high-spatial-resolution monitoring observation in the optical and UV with the \hst\ will also be important as the Doppler boosting and precession models make distinct predictions on the quiescent disc variability amplitude as a function of wavelength. Moreover, if the weak radio flux in GSN~069, $\simeq 47$~$\mu$Jy at $\sim 6$~GHz \citep{2019Natur.573..381M}, is associated with a compact jet, high-cadence radio observations with sufficient sensitivity might reveal variability consistent with orbital motion or a precessing jet. 

If the impacts plus modulation model is indeed correct, the X-ray QPEs detected in December 2018 in GSN~069 represent the first electromagnetic detection of an extragalactic, short period EMRI system, potentially opening the way to future electromagnetic and gravitational wave synergies and multi-messenger astronomy. As shown here, the QPEs properties and timing encode unique information of the likely complex galactic nuclei and SMBH inner environment, and can help understanding the structure and dynamics of accretion flows around recently activated SMBHs or, possibly, contribute to reveal the presence of hierarchical triples and tight, sub-milliparsec SMBH binaries in galactic nuclei.

\begin{acknowledgements}

 GM thanks Ari Laor for many valuable discussions on O-C data interpretation and Eric Agol for suggesting to use O-C diagrams in QPEs data analysis years ago. We also thank Chris Done for providing a version of the {\texttt optxagnf} model including retrograde accretion (negative black hole spin). We thank the \xmm, \cha, \swift, and \nicer\ teams for their excellent work and continued support. This work is based on observations obtained with \xmm, an ESA science mission with instruments and contributions directly funded by ESA Member States and NASA. We also used data obtained from the \cha\ X-ray mission, and software provided by the \cha\ X-ray Center (CXC). We acknowledge the use of public data from the \textit{Neil Gehrels \swift\ Observatory} data archive. This work made use of data supplied by the UK \swift\ Science Data Centre at the University of Leicester. We acknowledge the use of data and software provided by the High Energy Astrophysics Science Archive Research Center (HEASARC), which is a service of the Astrophysics Science Division at NASA/GSFC. GM acknowledges support by grant PID2020-115325GB-C31 funded by MCIN/AEI/10.13039/50110001103. AF acknowledges support provided by the "GW-learn" grant agreement CRSII5 213497 and the Tomalla Foundation. MB acknowledges support provided by MUR under grant ``PNRR - Missione 4 Istruzione e Ricerca - Componente 2 Dalla Ricerca all'Impresa - Investimento 1.2 Finanziamento di progetti presentati da giovani ricercatori ID:SOE\_0163'' and by University of Milano-Bicocca under grant ``2022-NAZ-0482/B''. MG is supported by the ``Programa de Atracci\'on de Talento'' of the Comunidad de Madrid, grant number 2022-5A/TIC-24235, and by Spanish MICIU/AEI/10.13039/501100011033 grant PID2019-107061GB-C61. RA was supported by NASA through the NASA Hubble Fellowship grant n. HST-HF2-51499.001-A awarded by the Space Telescope Science Institute, which is operated by the Association of Universities for Research in Astronomy, Incorporated, under NASA contract NAS5-26555. Support for this work was provided by NASA through the NASA Hubble Fellowship grant HST-HF2-51534.001-A awarded by the Space Telescope Science Institute, which is operated by the Association of Universities for Research in Astronomy, Incorporated, under NASA contract NAS5-26555. IL acknowledges support from a Rothschild Fellowship and The Gruber Foundation, as well as Simons Investigator grant 827103. AS acknowledges the financial support provided under the European Union’s H2020 ERC Consolidator Grant ``Binary Massive Black Hole Astrophysics'' (B Massive, Grant Agreement: 818691). 

\end{acknowledgements}

\bibliographystyle{aa}
\bibliography{biblio}

\begin{appendix}

\section{X-ray observations}
\label{sec:data}

In this work, we used X-ray data from observation performed with the \xmm, \cha, Niel Gehrels \swift, and \nicer\ X-ray observatories performed between December 2018 and September 2024. A summary of the observations used here is given in Table~\ref{tab:obs}. Below, we give some details on the data analysis that was performed for the different X-ray observatories and detectors we used.

\subsection{\xmm\ EPIC-pn and \cha\ ACIS-S}

\xmm\ EPIC-pn and \cha\ ACIS-S source and background products were extracted from circular regions on the same detector chip using the latest versions of the \texttt{SAS} (\xmm) and \texttt{CIAO} (\cha) dedicated software as well as the latest calibration. X-ray light curves were background subtracted, as well as corrected for various effects (bad pixels, quantum efficiency, vignetting, dead time) using the \texttt{epiclccorr} and \texttt{dmextract} tasks for \xmm\ and \cha\ respectively.  Although generally only a minor effect, the photon arrival times from all observations were barycentre-corrected in the DE405-ICRS reference system. This is generally irrelevant  except when deriving QPEs arrival times for the O-C analysis.

QPEs peak times where derived following \citet{2023A&A...670A..93M} from constant plus Gaussian functions fits to the individual light curves. We assume that the unknown time delay between an impact and the corresponding QPE peak is impact-independent, so that the QPE peak times can be taken as representative of impact times. This likely introduces a systematic uncertainty on the estimated impact times. Another option, followed for instance by \citet{2024PhRvD.109j3031Z,2024PhRvD.110h3019Z}, would be to consider as representative the start times of QPEs defined as the time when the observed count rate is a given fraction of the peak one. However, such a definition depends on the underlying quiescent count rate that might shift the QPE start times depending on the actual baseline count rate. QPEs peak times and duration have also been shown to be energy-dependent \citep{2019Natur.573..381M}. As the actual physical mechanism responsible for the X-ray emission is still uncertain, the peak time of QPEs in any given X-ray energy band can be considered as representative of the impact's time only if all QPEs share the same energy-dependent evolution from impact to peak. This is likely, but not guaranteed and introduces a possible further source of systematic error.  

In order to, at least partially, account for these systematic uncertainties and to reduce the risk of over-interpreting the data, we assign to each QPE peak time an uncertainty equal to half the time bin of the corresponding X-ray light curve (or uncertainties of $100$~s and $250$~s for \xmm\ and \cha\ observations). This is generally a factor of $2$-$3$ larger than the statistical-only $1\sigma$ uncertainty from actual constant plus Gaussian function fits, but we adopt this more conservative choice for the reasons expressed above. 

\begin{figure}[t] 
    \centering
        \includegraphics[width=0.9\columnwidth]{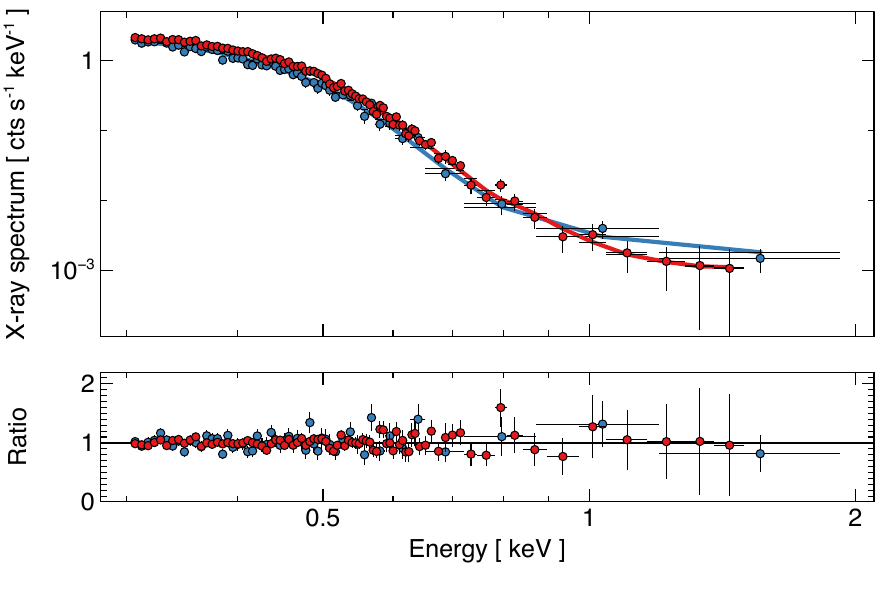}
    \caption{\xmm\ X-ray spectra from the 2024 campaign EPIC-pn X-ray spectra, best-fitting models, and data-to-model ratio for the first (red) and fourth (blue) \xmm\ observation of the 2024 campaign.}
    \label{fig:xmm24spec}
\end{figure}

\subsubsection{X-ray spectral analysis of the 2024 \xmm\ campaign}

GSN~069 was observed eight times by \xmm\ in 2024 to complement to X-ray monitoring campaign by \swift\ and \nicer. We are here interested in deriving the 0.3-1~keV flux from these observations and to compare them with those from \swift\ and \nicer\ that are shown in Fig.~\ref{fig:monitoring}. In order to obtain these measurements, we have obtained EPIC-pn X-ray spectra from all observations, extracting source and background products from circular regions on the same detector chip. No QPEs were detected during the 2024 observations, so that the full exposure was used to accumulated the X-ray spectra, excluding however periods of particularly high background (typically at the beginning or end of exposures). Appropriate redistribution matrices and ancillary files were generated for each observations by making use of the the {\texttt rmfgen} and {\texttt arfgen} tasks of the dedicated {\texttt SAS} software. 

The resulting X-ray spectra were all grouped to a minimum of $30$ background subtracted counts per bin and were fitted above 0.3~keV and up to the observation-dependent energy at which the signal was lost into the background. The spectra were analysed  within the {\texttt XSPEC} spectral analysis package \citep{1996ASPC..101...17A}, using $\chi^2$ minimisation. As discussed in previous works \citep{2023A&A...670A..93M}, the quiescent X-ray spectrum of GSN~069 is well described by a thermal disc model with typical temperature of $50$-$60$~eV and a weak power law component sometimes emerging above the background level above $\sim 1$-$2$~keV. We adopt a phenomenological power law continuum for the high-energy component and, since the photon index cannot be constrained reliably from the data, we fixed it to a common value of $\Gamma = 1.8$. 

\begin{table*}[h!]
        \centering
        \caption{Summary of observations.}
        \label{tab:obs}
        \begin{tabular}{lcccc }% four columns, alignment for each                                      
          \hline
\T  Mission & ObsID & Date (start) & Approx. exposure & QPEs \B \\
\hline
 \\ \multicolumn{5}{c}{Dec 2018 - May 2019 (O-C analysis) } \B \\ 
\hline
\T XMM & 0823680101 & 2018-12-24 & 60  &\textcolor{green}{\checkmark}   \B \\
\T XMM & 0831790701 & 2019-01-16 & 130 &\textcolor{green}{\checkmark}  \B \\
\T Chandra & 22096 & 2019-02-14    & 75  &\textcolor{green}{\checkmark} \B \\
\T XMM & 0851180401 & 2019-05-31 & 130 &\textcolor{green}{\checkmark} \B \\
\hline
\\ \multicolumn{5}{c}{2023 \xmm\ campaign } \B \\
\hline
\T XMM & 0914792701 & 2023-05-29 & 70 &\textcolor{green}{\checkmark} \B \\
\T XMM & 0914792901 & 2023-06-09 & 120 &\textcolor{green}{\checkmark} \B \\
\T XMM & 0914793101 & 2023-07-09 & 105 &\textcolor{green}{\checkmark} \B \\
\hline
\\ \multicolumn{5}{c}{2024 \xmm\ campaign$^a$ } \B \\
\hline
\T XMM (8) & 0943150101 & 2024-06-20 & 125 &\textcolor{red}{$\times$} \B \\
\T XMM (8) & ... & ... & ... &\textcolor{red}{$\times$} \B \\
\T XMM (8) & 0952390701 & 2024-07-18 & 30 &\textcolor{red}{$\times$} \B \\
\hline
\\ \multicolumn{5}{c}{2024 \swift\ campaign$^{a}$ } \B \\
\hline
\T \swift\ (23)& 00097778002 & 2024-05-15 & 2.4 &\textcolor{red}{$\times$} \B \\
\T \swift\ (23)& ... & ... & ... &\textcolor{red}{$\times$} \B \\
\T \swift\ (23)& 00010450054 & 2024-09-09 & 1.8 &\textcolor{red}{$\times$} \B \\
\hline
\\ \multicolumn{5}{c}{2024 \nicer\ campaign$^{a}$ } \B \\
\hline
\T \nicer\ (8)& 7672010401 & 2024-05-17 & 1.6 &\textcolor{red}{$\times$} \B \\
\T \nicer\ (8)& ... & ... & ... &\textcolor{red}{$\times$} \B \\
\T \nicer\ (8)& 7672011101 & 2024-08-25 & 3.6 &\textcolor{red}{$\times$} \B \\
\hline
\end{tabular}
\tablefoot{The last column indicates whether unambiguous QPEs were detected during the exposure (note that in many cases the exposures were too short to secure QPE detection). \\ $^a$ For the 2024 monitoring campaigns by \xmm, \swift, and \nicer, we only report the first and last monitoring observation used here, while the number of monitoring observations is reported in parenthesis in the first column. 
}
\end{table*}

The continuum is affected by ionised absorption associated with outflows, as discussed in detail by \citet{2024arXiv240617105K} who studied high-quality RGS data from all available \xmm\ observations of GSN~069. However, at the EPIC-pn camera's spectral resolution, and with typical exposures of $30$~ks, the spectral features associated with the ionised outflows are not prominent, and difficult to constrain. Using different absorption models can significantly affect the estimated intrinsic X-ray luminosity of the source, but has negligible effect on the flux measurements in which we are interested in here. We have nevertheless considered two different spectral models with either neutral or ionised intrinsic absorption (in excess of the fixed Galactic value with column density of $2.3\times 10^{20}$~cm$^{-2}$). For the ionised absorber, we use the {\texttt PION} photoionisation spectral model \citep{2016A&A...596A..65M} as in \citet{2024arXiv240617105K}. The spectral model, originally intended for use within the {\texttt SPEX} spectral analysis package \citep{1996uxsa.conf..411K}, has been converted into a table model for usage within {\texttt XSPEC} assuming that the slab is illuminated by an SED consistent with the thermal X-ray continuum in GSN~069. Based on the detailed analysis of high quality \xmm\ RGS data \citep{2024arXiv240617105K}, we impose a Nitrogen overabundance of $24$ with respect to Solar, in agreement also with previous studies based on \hst\ UV spectra \citep{2021ApJ...920L..25S}. After a few initial tests, we concluded that, as expected, the modest spectral resolution of the EPIC-pn camera did not allow to measure any outflow velocity reliably. This was then fixed to the value derived by \citet{2024arXiv240617105K}, namely $3\,000$~km~s$^{-1}$. The turbulent velocity was forced to be the same in all observations.

Both models produced a fair description of the data, but the ionised absorber model was statistically preferred ($\Delta\chi^2 = -38$ for a difference of $9$ in degrees of freedom), and resulted in $\chi^2 = 424$ for $420$ degrees of freedom. The ionised absorber column density was found to be in the range of $\sim 0.4$-$2\times 10^{22}$~cm$^{-2}$ with ionisation $\log\xi \simeq 3$-$4$, in broad agreement with the analysis presented in \citet{2024arXiv240617105K}. The derived 0.3-1~keV fluxes were then used as data points in Fig.~\ref{fig:xmm24}. In Fig.~\ref{fig:xmm24spec} we show the X-ray spectrum, best-fitting model, and resulting data-to-model ratio for two of the 2024 \xmm\ observations. In particular, we choose the first and fourth observations of the campaign as the former is the highest quality one, while the latter is representative of a low-flux state. In both cases, a high-energy power law (-like) continuum is well detected above $\sim 1$~keV, signalling that GSN~069 has an X-ray corona that up-scatters the softer disc photons to higher energies. As mentioned, the properties of the corona (electron temperature and optical depth) could not be constrained by the available data. 

\subsection{\swift\ XRT}

The \swift\ XRT observations were performed in Photon Counting (PC) mode and were analysed following the procedure outlined in \citet{2007A&A...469..379E,2009MNRAS.397.1177E}, which uses fully calibrated data and corrects for effects such as pile-up and the bad columns on the CCD, to obtain count rates on an observation-by-observation (or snapshot-by-snapshot) basis. Each observation typically comprises a couple of shorter exposure snapshots that have been also analysed following similar procedures to search for the occasional detection of one QPE (or part of it), although none was securely detected during the 2024 campaign. Specifically, the individual 0.3-1~keV count rates for every observation or snapshot exposure were obtained using the on-line \texttt{XRT product generator} tool\footnote{\href{https://www.swift.ac.uk/user\_objects}{\texttt{https://www.swift.ac.uk/user\_objects}}}.

\subsection{\nicer}

GSN 069 was observed by the Neutron Star Interior Composition ExploreR (\nicer; \citealt{2016SPIE.9905E..1HG}) for a total of 28~ks (PI: Miniutti, OBSIDs 7672010401-7672011101) from 17 May to 25 August 2024. We processed the data using \texttt{HEASoft} v6.33 and \texttt{NICERDAS} v11a. The quiescent flux of GSN 069 ($\sim 4-6\times 10^{-13}$ erg cm$^{-2}$ s$^{-1}$) is comparable to the \nicer\ observational background, which itself can vary significantly between snapshots. To reliably estimate the source flux over time, we followed the time-resolved spectroscopy procedure described in detail in \cite{2024ApJ...965...12C}, of which we give a brief summary here. We use \texttt{nimaketime} with unrestricted undershoot and overshoot event rates, and disabled event auto-screening on a per-focal plane module (FPM) basis; the typical resulting Good-Time Intervals (``GTIs'') lasted between 100-500~s, such that each OBSID was comprised of several GTIs spaced by $\mathcal{O}$(few~ks). In each GTI, we manually discard FPMs with 0-0.2 or 5-15 keV count rates $>3\sigma$ higher than the average, or above an absolute threshold of 20 counts sec$^{-1}$; such events are likely due to extreme background contamination via the solar wind charge exchange or cosmic rays, rather than being related to intrinsic source variability. We then used the \texttt{SCORPEON}\footnote{\href{https://heasarc.gsfc.nasa.gov/lheasoft/ftools/headas/niscorpeon.html}{\texttt{https://heasarc.gsfc.nasa.gov/lheasoft/ftools/\\headas/niscorpeon.html}}} template-based background model, together with an additional source model represented by \texttt{tbabs}$\times$\texttt{zashift}$\times$\texttt{diskbb}, to jointly fit for the source- and background-contribution to each GTI in {\texttt XSPEC}. The resulting source fluxes were estimated using the \texttt{cflux} convolutional model component on the source model, and associated errors are 1-sigma. Fluxes over one observation (typically within a few hours) were then combined, and the mean and standard deviation is shown in Fig.~\ref{fig:monitoring} on an observation-by-observation basis. 

\section{Further O-C analysis in GSN~069}
\label{sec:furtherOC}

As discussed in Section~\ref{sec:gsnOC}, the QPE identification used to produce the O-C diagrams in Fig.~\ref{fig:GSNOCmix} is not unique, and we report below results obtained with two different identifications for QPEs during the May 2019 observation that is the most uncertain as it is associated with the longest gap with respect to the previous one ($\sim 107$~d, the second longest being $\sim 29$~d). 

\subsection{Alternative QPE identification in May 2019}
\label{sec:GSNOCdiffID}

\begin{figure*}[t] 
    \centering
        \includegraphics[width=0.45\textwidth]{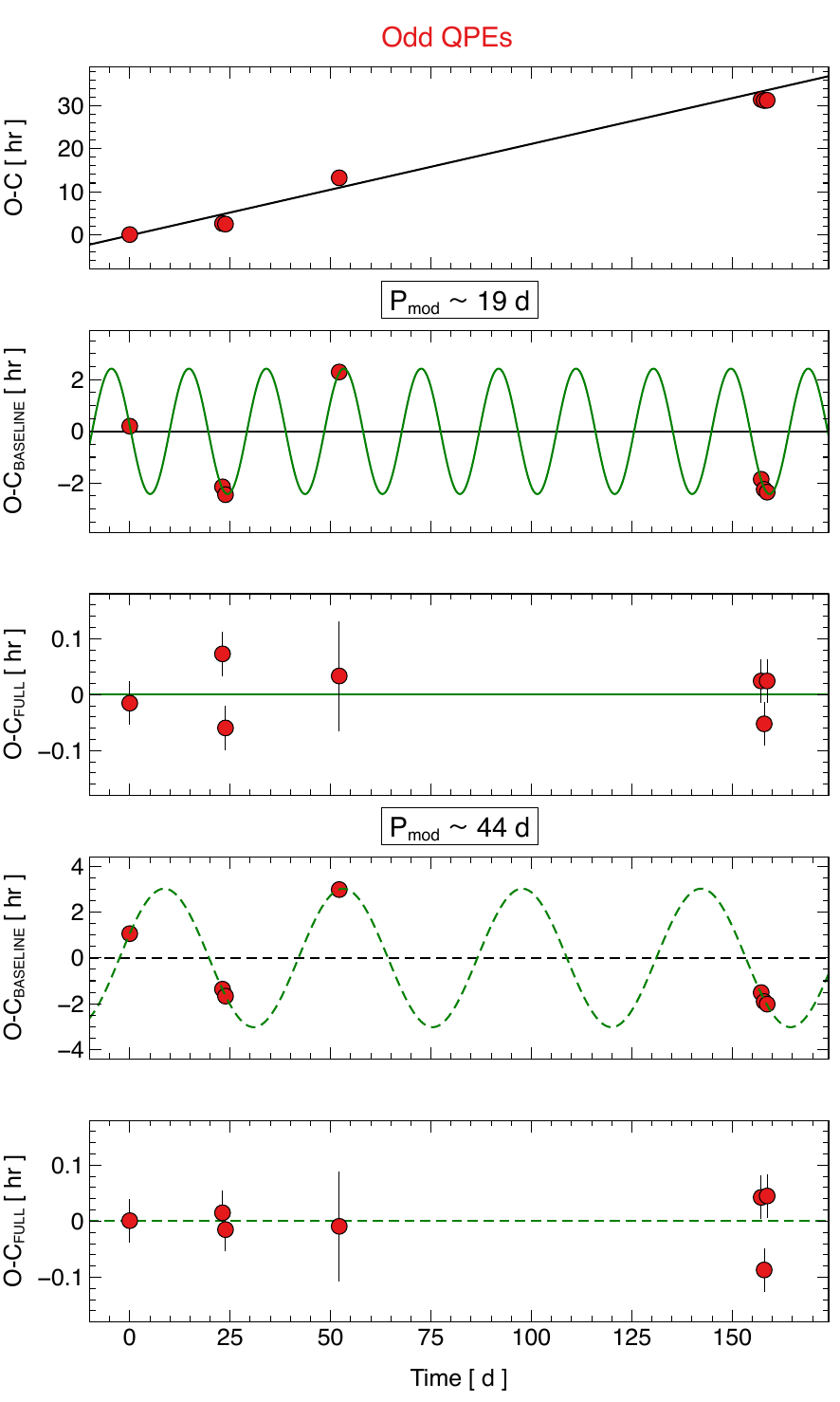}%
    \hspace{0.3cm}%
        \includegraphics[width=0.45\textwidth]{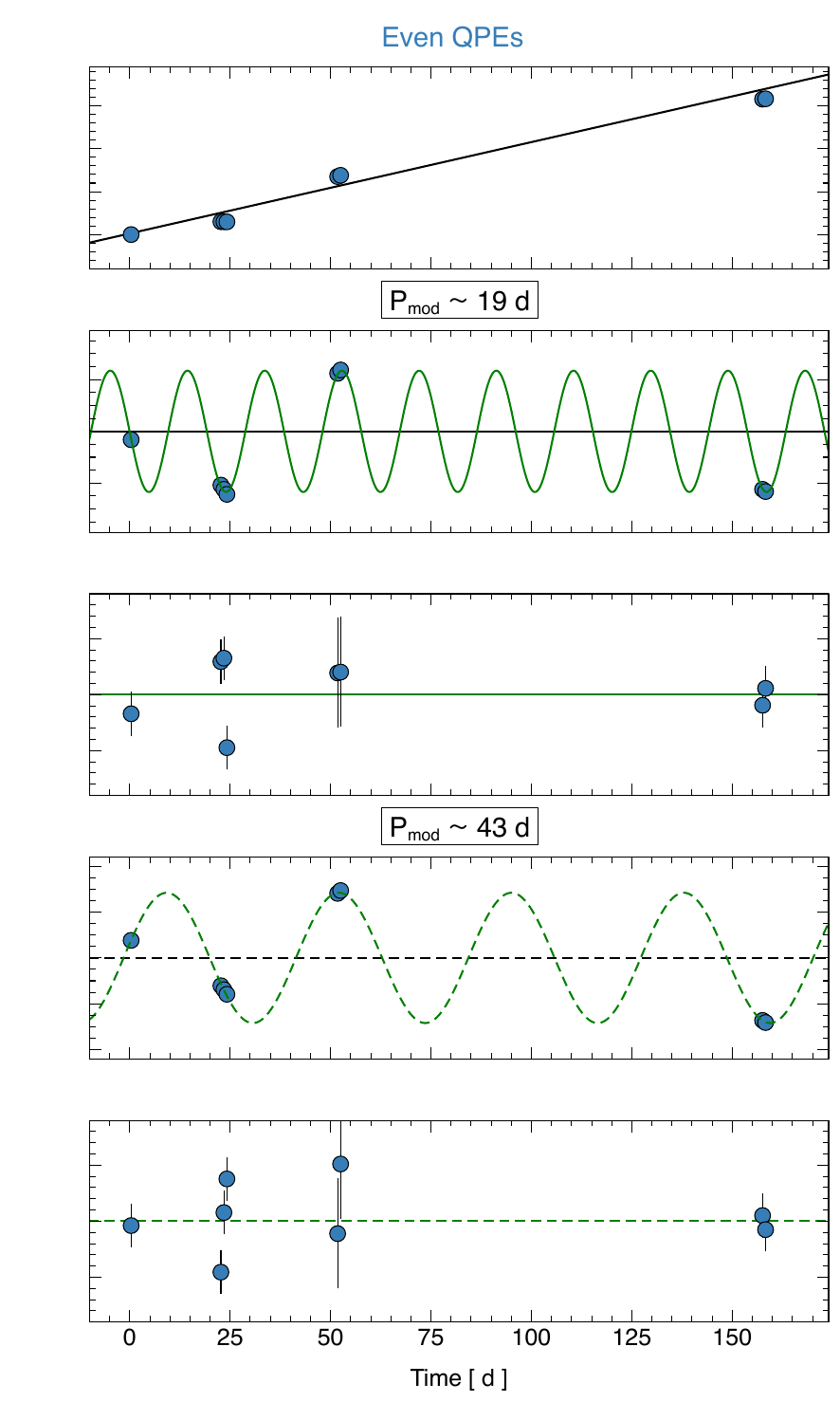}%
    \caption{O-C diagrams for GSN~069: linear baseline model. Same as Fig.~\ref{fig:GSNOCmix}, but the first QPE in May 2019 is here identified with the 211th odd QPE.}
    \label{fig:GSNOClin}
\end{figure*}

\begin{figure*}[t] 
    \centering
        \includegraphics[width=0.45\textwidth]{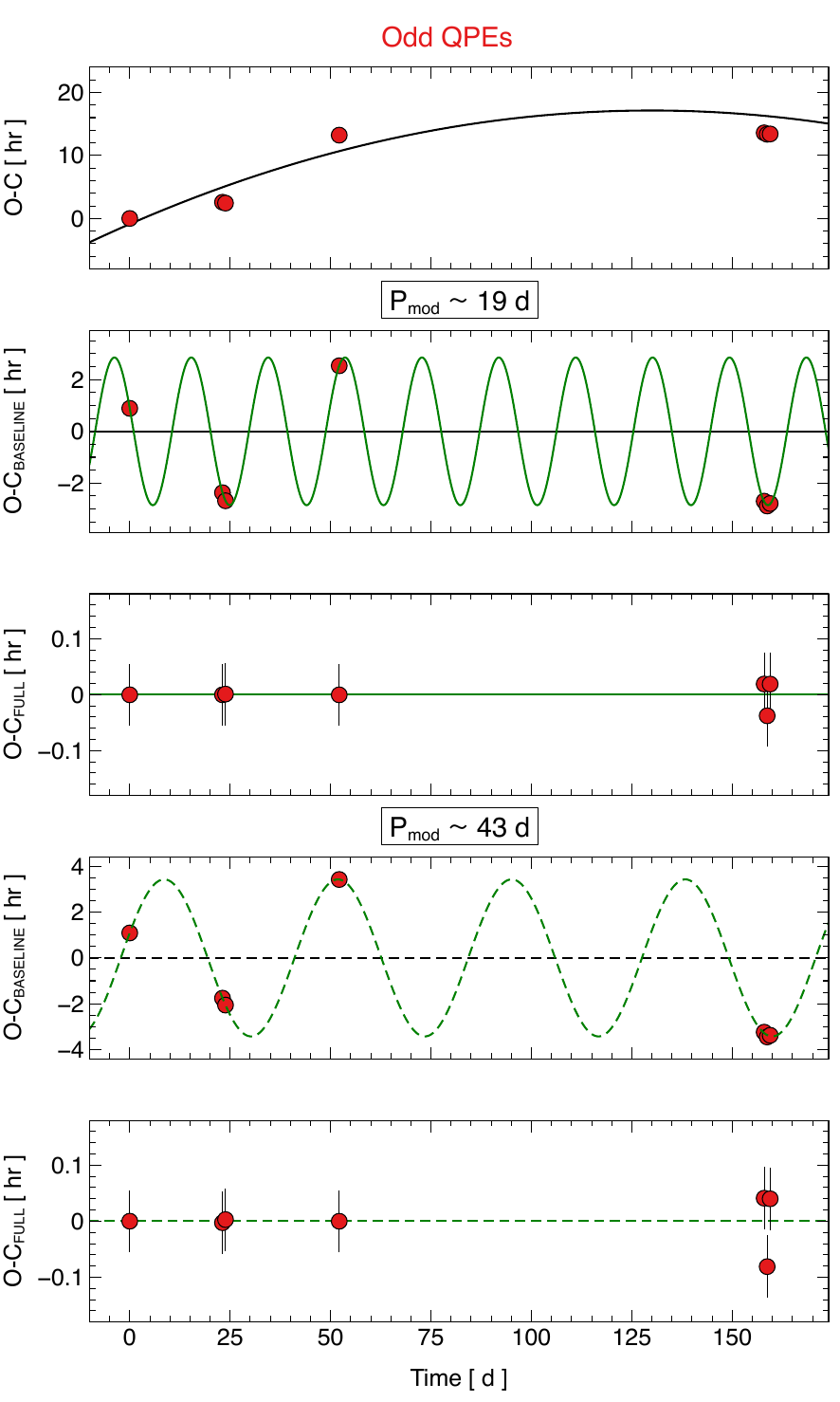}%
    \hspace{0.3cm}%
        \includegraphics[width=0.45\textwidth]{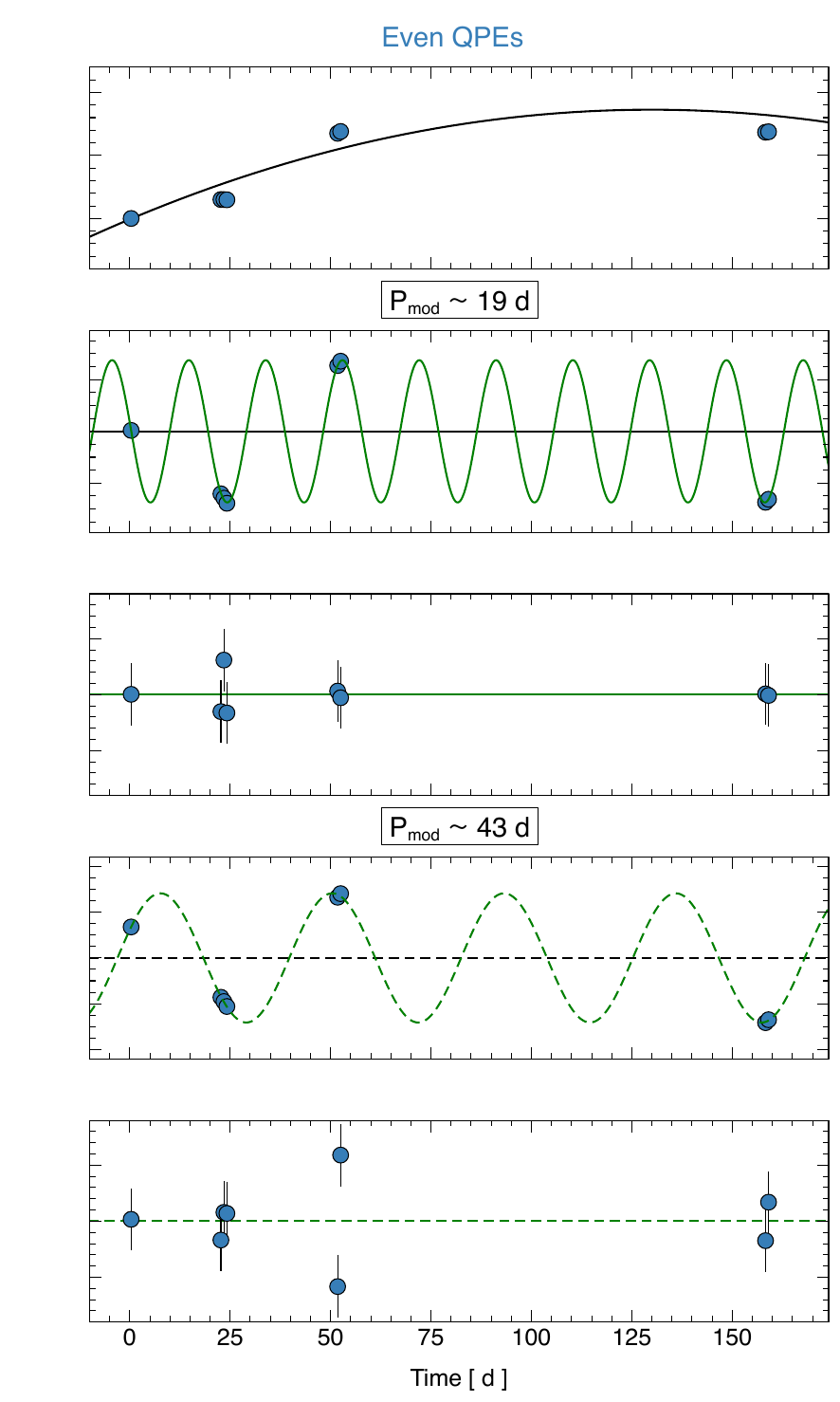}%
    \caption{O-C diagrams for GSN~069: parabolic baseline model. Same as Fig.~\ref{fig:GSNOCmix}, but the first QPE in May 2019 is here identified with the 212th odd QPE.}
    \label{fig:GSNOCquad}
\end{figure*}

In the original version of the O-C analysis in Section~\ref{sec:gsnOC}, the first QPE in May 2019 was assumed to be the 211th even one. The two closest versions of O-C diagrams for GSN~069 are obtained by assuming that it is instead either the 211th or the 212th odd one (the identification of all other QPEs follows trivially). The net result is that all O-C data for the May 2019 observation shift upwards or downwards by half $P_{\rm trial}$ with respect to the case shown in Fig.~\ref{fig:GSNOCmix}. The O-C diagrams for these two different QPE identifications are shown in Fig.~\ref{fig:GSNOClin} and \ref{fig:GSNOCquad} respectively. 

The O-C diagrams in Fig.~\ref{fig:GSNOClin} are well described by a linear baseline model (solid line in the upper panels) plus a sinusoidal modulation, while no parabolic trend is needed so that $\dot{P}_{\rm orb} = 0$. On the other hand, the O-C diagrams for the alternative QPE identification shown in Fig.~\ref{fig:GSNOCquad} do require an additional  parabolic trend (solid line in the upper panels) and are therefore associated with $\dot{P}_{\rm orb}\neq 0$. As was the case for the O-C analysis in Section~\ref{sec:gsnOC}, a periodic modulation with either a $\sim 19$~d or $\sim 43$-$44$~d period is required in both cases. Results are reported in Table~\ref{tab:OCgsnlin} and Table~\ref{tab:OCgsnquad} for the $\dot{P}_{\rm orb} = 0$ and the $\dot{P}_{\rm orb} \neq 0$ solutions respectively. 

The orbital period $P_{\rm orb}$ is always the same, within errors, for both odd and even QPEs regardless of $P_{\rm mod}$ and of the actual QPE identification, and it is also consistent with that obtained in Section~\ref{sec:gsnOC}. All other parameters (except $\dot{P}_{\rm orb}$, as obvious) are also consistent with the previous analysis at the few per cent level. The only significant difference between the three versions of the O-C diagrams we have produced is in the derived EMRI's orbital evolution. We suggest three possible values associated with the three different identification of QPEs in May 2019, namely $\dot{P}_{\rm orb} =0$, $-3$-$4\times 10^{-5}$, or $-6$-$7\times 10^{-5}$. As discussed in Section~\ref{sec:Pdotconfirm}, the 2023 campaign on GSN~069 suggests that the intermediate value might be correct, although we consider the detection of $\dot{P}_{\rm orb} \simeq -3.7\times 10^{-5}$ as tentative only due to the small number of independent measurements during the 2023 \xmm\ campaign. The stability of best-fitting parameters despite the difference in QPE identification provides support to the robustness of the overall O-C analysis and derived parameters.

\begin{table}[t]
        \centering
        \caption{O-C analysis for GSN~069 leading to $\dot{P}=0$. Summary of results from the O-C analysis of QPEs in GSN~069 assuming that the first QPE in May 2019 is the 211th odd one. The corresponding O-C diagrams are shown in Fig.~\ref{fig:GSNOClin}. }
        \label{tab:OCgsnlin}
        \begin{tabular}{lcc} % four columns, alignment for each                                      
          \hline
\T  & Odd &  Even  \B \\
\hline          
\T $P_{\rm mod} \sim 19$~d & &   \B \\
\hline
\T $P_{\rm orb}$~[hr] & $18.04\pm 0.04$ & $18.04\pm 0.04$ \B \\
\T $P_{\rm mod}$~[d] & $19.27\pm 0.05$ & $19.22\pm 0.06$ \B \\
\T A$_{\rm mod}$~[hr] & $2.42\pm 0.06$ & $2.35\pm 0.06$ \B\\
\hline
\T $P_{\rm mod} \sim 43$-$44$~d & &  \B \\
\hline
\T $P_{\rm orb}$~[hr] & $18.04\pm 0.04$ & $18.05\pm 0.04$ \B \\
\T $P_{\rm mod}$~[d] & $44.5\pm 0.6$ & $43.1\pm 0.7$ \B \\
\T A$_{\rm mod}$~[hr] & $3.0\pm 0.1$ & $2.85\pm 0.09$ \B \\
\hline
        \end{tabular}
\end{table}

\begin{table}[t]
        \centering
        \caption{O-C analysis for GSN~069 leading to $\dot{P}\neq 0$. Summary of results from the O-C analysis of QPEs in GSN~069 assuming that the first QPE in May 2019 is the 212th odd one. The corresponding O-C diagrams are shown in Fig.~\ref{fig:GSNOCquad}. }
        \label{tab:OCgsnquad}
        \begin{tabular}{lcc} % four columns, alignment for each                                      
          \hline
\T  & Odd &  Even  \B \\
\hline          
\T $P_{\rm mod} \sim 19$~d & &   \B \\
\hline
\T $P_{\rm orb}$~[hr] & $18.08\pm 0.04$ & $18.07\pm 0.04$ \B \\
\T $\dot{P}_{\rm orb}$~[$10^{-5}$] & $-6.5\pm 0.8$ & $-6.3\pm 0.7$ \B \\
\T $P_{\rm mod}$~[d] & $19.1\pm 0.1$ & $19.2\pm 0.1$ \B \\
\T A$_{\rm mod}$~[hr] & $2.86\pm 0.07$ & $2.75\pm 0.08$ \B\\
\hline
\T $P_{\rm mod} \sim 43$~d & &  \B \\
\hline
\T $P_{\rm orb}$~[hr] & $18.07\pm 0.04$ & $18.08\pm 0.04$ \B \\
\T $\dot{P}_{\rm orb}$~[$10^{-5}$] & $-5.5\pm 1.0$ & $-7.2\pm 0.9$ \B \\
\T $P_{\rm mod}$~[d] & $43.2\pm 0.5$ & $42.8\pm 0.7$ \B \\
\T A$_{\rm mod}$~[hr] & $3.4\pm 0.6$ & $2.8\pm 0.4$ \B \\
\hline
        \end{tabular}
\end{table}

\subsection{Fitting procedure for the O-C data of Fig.~\ref{fig:GSNOCmix}}
\label{sec:fitprocedure}

As mentioned in Section~\ref{sec:gsnOC} , the number of data points in the odd QPEs (six) time series in Fig.~\ref{fig:GSNOCmix} is the same that of the  free parameters of the adopted model $a+bx+cx^2+A_{\rm mod}\sin(P_{\rm mod},\phi_{\rm mod})$ prevented us to formally apply the model in that case. We then first considered fits to the even QPEs data (nine data points) and derived the best-fitting parameters and uncertainties.  We then fitted the odd QPEs data by fixing one of model's parameters to the best-fitting value derived from the even QPEs time series. The procedure was repeated by changing the value of the fixed parameter exploring a range of possible values much larger than the corresponding uncertainty from the even QPEs data best-fitting results. Specifically, we chose $P_{\rm orb}$ because this is expected to be the same for both odd and even QPEs, and explored $17 \leq P_{\rm orb} \leq 19 $~hr in steps of $0.01$~hr in the odd QPEs data. Choosing a different parameter did not produce any significant change in the results. With this procedure, we could estimate the allowed range for all parameters and compare them with results from the even QPEs time series. Results are reported in Table~\ref{tab:OCgsnmix}. The procedure turned out to be validated by the excellent agreement with the independent analysis of the even QPEs time series for all parameters, as well as with results discussed in Section~\ref{sec:GSNOCdiffID}, where independent fits on both odd and even QPEs data could be performed.

\FloatBarrier

\section{Comments on SMBH spin effects and numerical implementation}
\label{sec:notesmodel}

In this work, we made use of the numerical code presented in \citet{2023A&A...675A.100F} who implemented a semi-analytic approach to numerically simulate the collisions between the accretion disc around a SMBH with mass $M$ and the secondary orbiting companion with mass $m\ll M$ in an EMRI system. The accretion disc around the primary was assumed to extend down to the innermost stable circular orbit of a generic Kerr black hole, and to be prograde with respect to the SMBH spin that was assumed to be aligned with the $z$-axis (i.e. the disc's angular momentum has also positive $z$-component). Based on the growing evidence for a connection between QPE systems and TDEs, the considered disc structure was assumed to be consistent with a TDE origin following earlier work by \citet{2016MNRAS.455.1946F}. The equations of motion of the EMRI were integrated up to the 3.5 post-Newtonian order, thus removing the test-particle approximation used in previous work \citep[e.g.][]{2021ApJ...921L..32X}, and including all relevant time delays (Roemer, Shapiro, and Einstein delays) from the impact position to the observer. Fir further details, we refer to the work by \citet{2023A&A...675A.100F}.

Rigid disc precession can be naturally implemented in the numerical code. Whenever the primary SMBH is spinning, and the disc misaligned, we assumed that the warp induced by the Lense-Thirring effect \citep{1918PhyZ...19..156L} propagates as a bending wave, allowing the disc to rigidly precess around the primary, spinning SMBH. In this regime, the disc precesses rigidly with a frequency that is the angular-momentum-weighted average of the Lense-Thirring precession frequency over the extent of the disc. In our work, however, to study the effects of disc precession alone without introducing unnecessary complications due to the EMRI orbit's nodal precession induced by the spinning SMBH (see discussion below), we have included disc precession at an arbitrarily chosen period, keeping the SMBH black hole spin fixed to zero.

\begin{figure}[t]
\centering \includegraphics[width=0.9\columnwidth]{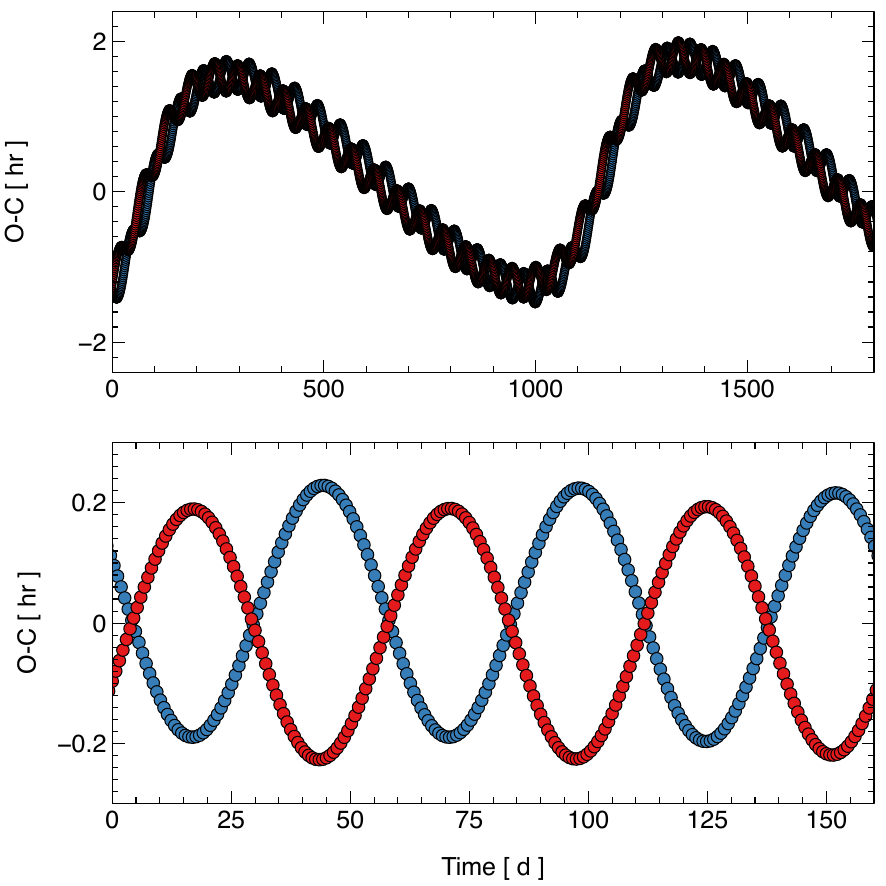}
\caption{Primary SMBH spin effects: nodal precession. O-C diagrams for the same parameters as in Fig.~\ref{fig:ex}, but where we have considered an almost maximally-rotating Kerr primary SMBH with dimensionless spin of $\chi =0.9$. The upper panel shows the O-C diagrams over slightly more than one nodal precession period ($\sim 3$~yr), while the lower panel are O-C diagrams derived from a restricted $160$~d portion of the full light curve, commensurate with the baseline that is probed in GSN~069 by observations between December 2018 and May 2019 (see Fig.~\ref{fig:gsn} and \ref{fig:GSNOCmix}).}
\label{fig:spin}
\end{figure}

Whenever the SMBH is spinning, and even in absence of disc precession, the EMRI orbit is affected because, in addition to the in-plane apsidal precession that is present also for a non-spinning SMBH, it is subject to nodal precession of the orbital plane at the nodal precession frequency. Nodal precession affects the QPEs timing properties in a similar way as disc precession, introducing a coherent modulation of QPEs arrival times that induces correlated O-C diagrams for odd and even QPEs at the nodal precession timescale. However, the nodal precession timescale is generally much longer than the disc precession and apsidal precession ones, so that its effects are likely to be important only on very long timescales, typically much longer than the baseline spanned by QPEs observations. The effect of nodal precession on the O-C diagrams is shown in Fig.~\ref{fig:spin} where we assumed exactly the same parameters as those resulting in Fig.~\ref{fig:ex}, but considered the case of an almost maximally-spinning Kerr SMBH with dimensionless spin $\chi =0.9$. In the upper panel of Fig.~\ref{fig:spin}, we show the O-C diagrams on a relatively long baseline covering slightly more than one nodal precession timescale. The precession of the EMRI orbital plane introduces a long-term correlation between the O-C diagrams of odd and even QPEs on the nodal precession timescale. However, on a baseline only comprising a few apsidal precession timescales, the O-C diagrams for the two branches are still nearly perfectly anti-correlated as shown in the lower panel, where we selected randomly a $160$~d time interval from the simulation to mimic the baseline of GSN~069 observations (see Fig.~\ref{fig:gsn} and, e.g., \ref{fig:GSNOCmix}), and repeated the O-C analysis.

As mentioned, the nodal precession timescale is always significantly longer than the apsidal one ($\sim 3$~yr versus $\sim 50$~d for the case shown in Fig~\ref{fig:spin}), so that black hole spin effects can only be revealed by very long-term, high-cadence monitoring campaigns. As we are here typically interested with observations spanning at most a few apsidal precession timescales, we do not discuss nodal precession of the EMRI orbital plane. However, its effects may be needed to interpret the long term-behaviour of at least some QPE sources in the future. 

The simulations involving a hierarchical triplet have been performed leveraging on the computational framework developed in \citet{2016MNRAS.461.4419B,2018MNRAS.477.3910B}. Specifically, the code integrates the equations of motion of three-body systems exploiting the high accuracy Bulirsch-Stoer integration scheme. The numerically obtained trajectories were computed from Post-Newtonian three-body Hamiltonian featuring correction up to 2.5 PN order. The framework, by allowing generic mass ranges for the three bodies, can be initialised to evolve an EMRI system perturbed by a third object. In more details, the system was initialised with an inner binary $M_1-m$ and and outer binary effectively formed by $M_1+m$ and $M_2$. Here, $m << M_1 \sim M_2$. For the specific simulations of this work, the three-body code has been complemented with a routine to detect the crossing of a disc orbiting around $M_1$. Each time the trajectory of $m$ intercepted the plane defined by the accretion disc, its coordinates and the crossing time were recorded. The procedure closely follows the one employed in \citet{2023A&A...675A.100F}. Once the time of crossing was recorded, it was complemented by the appropriate time-delays (Roemer, Shapiro and Einstein) referred to a distant observer, as done in \citet{2023A&A...675A.100F}. 
Specifically, we expressed the arrival time to the distant observer as \citep[see e.g.][sec. 10.3.6]{2014grav.book.....P} as:
%%%%%%%%%%%%%%%%%%%%%%%
\begin{equation}
    t_a = \tau + \Delta_R(\tau) + \Delta_S(\tau) + \Delta_E(\tau),
\end{equation}
%%%%%%%%%%%%%%%%%%%%%%%
where the three delays are given by the following expression:
%%%%%%%%%%%%%%%%%%%%%%%
\begin{align}
    \Delta_R(\tau) &= |\mathbf{r}_{\rm obs}-\mathbf{r}_{2}|/c,\\
    \Delta_S(\tau) &= \frac{2GM_1}{c^3}\ln\left(\frac{|\mathbf{r}_{\rm obs}-\mathbf{r}_{1}| + (\mathbf{r}_{\rm obs}-\mathbf{r}_{1})\cdot \mathbf{k}}{r + \mathbf{r}\cdot\mathbf{k}}\right),\\
    \Delta_E(\tau) &= \frac{M_2+2M_1}{M}\sqrt{\frac{a^3}{G M}}\frac{G M_1}{a c^2} \ \sin(u).
\end{align}
%%%%%%%%%%%%%%%%%%%%%%%
Here, $M = M_1 + M_2 $ is the mass of the binary, while $\mathbf{r}_{\rm obs}$ denotes the observer position. $\mathbf{r}_{1}$ and $\mathbf{r}_{2}$ represent instead the position vectors of the primary and secondary SMBH with respect to the centre of mass of the system, while the unit vector $\mathbf{k}$ is given by:
%%%%%%%%%%%%%%%%%%%%%%%
\begin{equation}
    \mathbf{k} = \frac{\mathbf{r}_{\rm obs}-\mathbf{r}_{2}}{|\mathbf{r}_{\rm obs}-\mathbf{r}_{2}|}.
\end{equation}
%%%%%%%%%%%%%%%%%%%%%%%
Finally, $a$, $e$, and $u$ respectively represent the orbit semi-major axis, eccentricity and eccentric anomaly, respectively. The time series of (observed) crossing times, associated with QPEs, were then used to compute the O-C plots and all other relevant quantities.

\section{Schematic representation of the proposed modulations}
\label{sec:r1}

\begin{figure*}[h] 
    \centering
        \includegraphics[width=1.99\columnwidth]{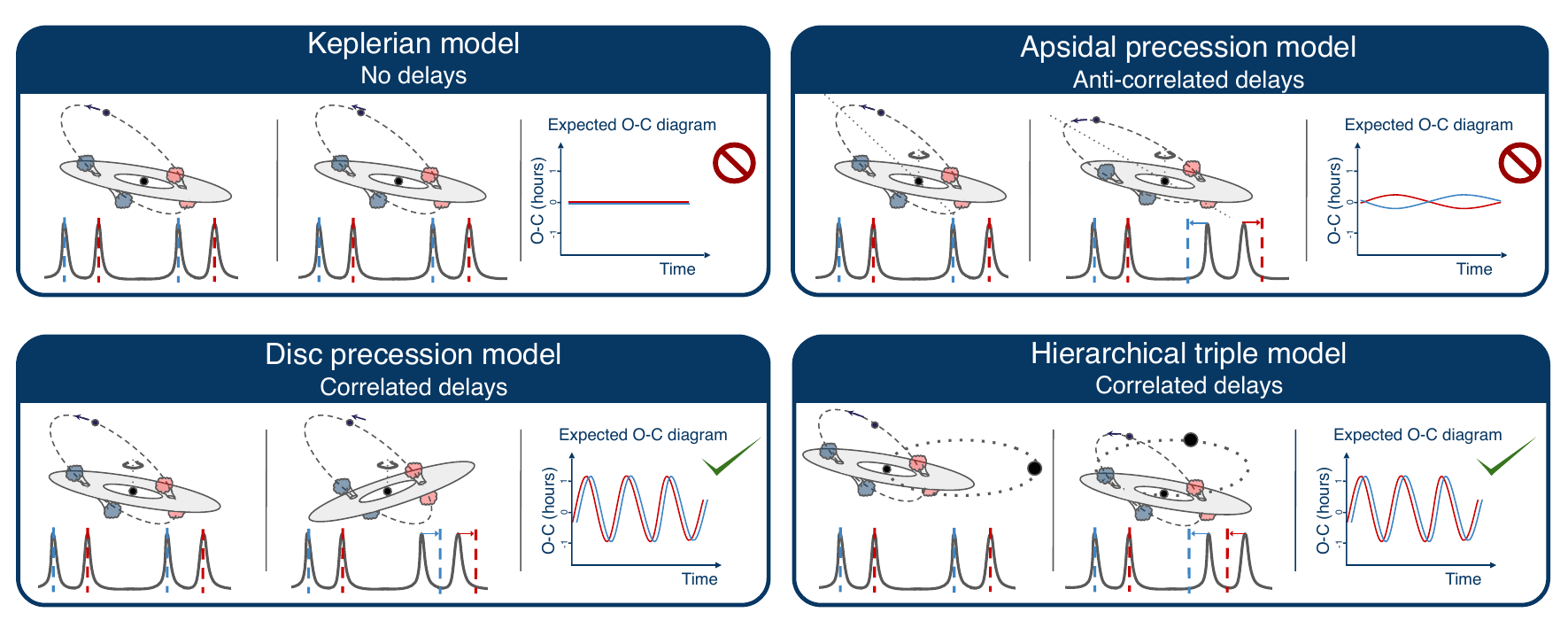}
\caption{Schematics of the effect of apsidal precession and external modulations on O-C diagrams. The upper panels show the case of a Keplerian EMRI orbit (left) and of a General Relativistic one including apsidal precession (right). The lower two panels are associated with the two modulating scenarios we considered in our work: disc precession (left) and a hierarchical triple model in which an outer SMBH forms a tight SMBH binary with the inner EMRI system (right), see text for details. }
    \label{fig:r2}
\end{figure*}

In Fig.~\ref{fig:r2}, we show a schematic representation of the effect of apsidal precession and of the proposed external modulations on O-C diagrams. At each impact, two-sided, hot and expanding plasma clouds are ejected from the disc, and are responsible for QPEs \citep{2023A&A...675A.100F,2023ApJ...957...34L}. The clouds colour-code distinguishes between impacts at the ascending and descending nodes that are associated with, e.g., odd and even QPEs. Each panel comprises three different sub-panels. The first two represent two different time snapshots of the system geometry and the resulting QPE light curve where the dashed blue and red lines indicate the QPE arrival times expected if QPEs of the same parity recur on a constant period, as is the case for a Keplerian EMRI orbit. The leftmost sub-panel shows the resulting O-C diagrams.

In the upper-left panel of Fig.~\ref{fig:r2}, we show the case of a Keplerian EMRI orbit. As the Keplerian orbit is fixed, impacts occur always at the same site on the disc, and consecutive odd (even) QPEs are always separated by the same time interval, corresponding to the EMRI orbital period. The O-C data for odd and even QPEs align on a straight line and, once a linear ephemeris is removed, they are actually all zero. The upper-right panel shows the effects of General Relativistic apsidal precession of the EMRI orbit. Impacts at the ascending nodes (odd QPEs) are delayed with respect to the Keplerian expectation, while the corresponding ones at the descending nodes (even QPEs) are anticipated by a similar time interval (or vice-versa). Therefore, the O-C diagrams for odd and even QPEs are modulated on the apsidal precession timescale and are in phase opposition. The amplitude of the modulation is of the order of few minutes for typical EMRI parameters and observer inclination. 

The lower panels of Fig.~\ref{fig:r2} show the effects of the two external modulation scenarios considered in our work, namely rigid disc precession (lower-left) or the presence of an outer SMBH forming a SMBH binary with the inner EMRI system (lower-right). In both cases, if the ascending impact is delayed (anticipated) with respect to the Keplerian expectation, the corresponding impacts at the descending node is also delayed (anticipated). The O-C diagrams for odd and even QPEs are therefore modulated on the external timescale (disc precession or outer SMBH binary orbital period) with minimal phase difference. A small phase difference is expected due to the fact that impacts at the ascending and descending nodes are not simultaneous but separated by the average recurrence time between an ascending node impact and the subsequent descending node one ($<T_{\rm rec}> \simeq 9$~hr in GSN~069).

\end{appendix}
 
\end{document}